\documentclass[sigconf,screen]{acmart}
\setcopyright{rightsretained}
\acmPrice{}
\acmDOI{10.1145/3611643.3616350}
\acmYear{2023}
\copyrightyear{2023}
\acmSubmissionID{fse23main-p1207-p}
\acmISBN{979-8-4007-0327-0/23/12}
\acmConference[ESEC/FSE '23]{Proceedings of the 31st ACM Joint European Software Engineering Conference and Symposium on the Foundations of Software Engineering}{December 3--9, 2023}{San Francisco, CA, USA}
\acmBooktitle{Proceedings of the 31st ACM Joint European Software Engineering Conference and Symposium on the Foundations of Software Engineering (ESEC/FSE '23), December 3--9, 2023, San Francisco, CA, USA}
\received{2023-02-02}
\received[accepted]{2023-07-27}

\usepackage{adjustbox}
\usepackage{booktabs}
\usepackage{xspace}
\usepackage{hyperref}
\usepackage{graphicx}
\usepackage{multirow}
\usepackage{listings}
\usepackage{makecell}
\usepackage{enumitem}
\usepackage{tikz}
\usepackage{subcaption}
\usepackage{caption}
\usepackage{balance}
\usepackage{microtype}

\definecolor{gray}{RGB}{211,211,211}
\definecolor{javared}{rgb}{0.6,0,0} %
\definecolor{javagreen}{rgb}{0.25,0.5,0.35} %
\definecolor{javapurple}{rgb}{0.5,0,0.2} %
\definecolor{javadocblue}{rgb}{0.25,0.35,0.75} %
\newcommand{\jbasicstyle}{\small\sffamily} %

\newcommand{\jnumberstyle}{\scriptsize}

\lstdefinelanguage{pseudo}
{
  morekeywords={},
  keywordstyle=\bfseries,
  lineskip=-0.1em,
  numbers=left, %
  numberstyle=\jnumberstyle,
  numbersep=4pt,
  basicstyle=\jbasicstyle,
  breaklines=true,
  breakautoindent=true,
  tabsize=2,
  columns=fullflexible,
  morecomment=*[l][\textsl]{//},
  mathescape=true,
  xleftmargin=10pt,
}

\lstdefinelanguage{todo-comment}
{
  morekeywords={},
  keywordstyle=\bfseries,
  lineskip=-0.1em,
  numbers=none,
  basicstyle=\jbasicstyle,
  breaklines=true,
  breakautoindent=true,
  tabsize=2,
  columns=fullflexible,
  morecomment=*[l][\textsl]{//},
  mathescape=true,
  xleftmargin=-10pt,
}

\lstdefinelanguage{java-pretty}
{
  language=java,
  basicstyle=\scriptsize\ttfamily,
  numberstyle=\scriptsize,
  breaklines=true,
  columns=fullflexible,
  showstringspaces=false,
  keywordstyle=\bfseries\color{javadocblue},
  stringstyle=\color{javared},
  commentstyle=\color{javagreen},
  numbersep=-12pt,
  morecomment=[s][\color{javadocblue}]{/**}{*/},
  moredelim=[is][\color{red!50!black}]{(-W<)}{(>W-)},
  moredelim=[is][\bfseries\color{red!80!black}]{(-w<)}{(>w-)},
  moredelim=[is][\color{green!30!black}]{(+W<)}{(>W+)},
  moredelim=[is][\bfseries\color{green!60!black}]{(+w<)}{(>w+)},
}

\usepackage[inkscapeformat=png]{svg}

\usetikzlibrary{shapes,arrows,positioning,calc}

\tikzstyle{text_box} = [rectangle, text width=15em, text centered]
\tikzstyle{block} = [draw=black, thick, rectangle, text=black, text width=8em, text centered, rounded corners, minimum height=7ex] %
\tikzstyle{line} = [draw, line width=0.25mm, -latex']
\tikzstyle{long_text_box} = [rectangle, text width=15em, minimum size=2cm]

\newcommand{\XComment}[1]{}

\newcommand{\DefMacro}[2]{\expandafter\newcommand\csname rmk-#1\endcsname{#2}}
\newcommand{\UseMacro}[1]{\csname rmk-#1\endcsname}
\newcommand{\RQ}[2]{\vspace{3pt}\noindent\textbf{RQ#1:} #2}

\newcommand{\MyPara}[1]{\vspace{2pt}\noindent\textbf{#1}.}

\newcommand{\InputWithSpace}[1]{\bgroup\def\arraystretch{1.3}\input{#1}\egroup}
\newcommand{\Code}[1]{{\ifmmode{\mathtt{#1}}\else$\mathtt{#1}$\fi}}
\newcommand{\CodeIn}[1]{{\ifmmode{\mathtt{#1}}\else$\mathtt{#1}$\fi}}

\usepackage{mdwlist}

\newcommand{\mycheckmark}{{\normalsize \checkmark}\xspace}
\newcommand{\mycross}{$\mathbin{\tikz [x=1.4ex,y=1.4ex,line width=.2ex] \draw (0,0) -- (1,1) (0,1) -- (1,0);}$\xspace\xspace}

\newenvironment{myquote}%
  {\list{}{\leftmargin=6pt\rightmargin=0pt}\item[]}%
  {\endlist}
\newcommand{\MyQuote}[1]{\begin{myquote} #1 \end{myquote}}

\newcommand{\Title}{Multilingual Code Co-Evolution Using Large Language Models}

\newcommand{\APIs}{APIs\xspace}

\newcommand{\LLM}{LLM\xspace}
\newcommand{\LLMs}{LLMs\xspace}

\newcommand{\DeltaTool}{\textsc{Codeditor}\xspace}
\newcommand{\Copy}{Copy\xspace}
\newcommand{\translationModel}{CodeT5-Translation\xspace}
\newcommand{\CopyEdits}{CopyEdits\xspace}
\newcommand{\edittranslation}{EditsTranslation\xspace}
\newcommand{\metaEdits}{MetaEdits\xspace}
\newcommand{\Codex}{Codex\xspace}
\newcommand{\chatgpt}{ChatGPT\xspace}
\DefMacro{CoditT5}{CoditT5\xspace}
\DefMacro{CodeT5Up}{CodeT5-Update\xspace}

\newcommand{\Train}{Train\xspace}
\newcommand{\Val}{Val\xspace}
\newcommand{\Test}{Test\xspace}

\newcommand{\CodeTf}{CodeT5\xspace}
\newcommand{\pretrained}{pretrained\xspace}
\newcommand{\pretraining}{pretraining\xspace}
\newcommand{\finetuning}{fine-tuning\xspace}
\newcommand{\finetuned}{fine-tuned\xspace}
\newcommand{\finetune}{fine-tune\xspace}
\newcommand{\Csharp}{C\#\xspace}

\newcommand{\deltatask}{multilingual co-editing\xspace}
\newcommand{\concise}{concise\xspace}  %
\newcommand{\Concise}{Concise\xspace}
\newcommand{\unambig}{unambiguous\xspace}
\newcommand{\Unambig}{Unambiguous\xspace}
\newcommand{\mode}{mode\xspace}
\newcommand{\modes}{modes\xspace}

\DefMacro{java2cs}{J2CS\xspace}
\DefMacro{cs2java}{CS2J\xspace}
\DefMacro{java2cs_cp}{J2CS-CP\xspace}
\DefMacro{cs2java_cp}{CS2J-CP\xspace}

\DefMacro{source}{\ensuremath{S}}
\DefMacro{target}{\ensuremath{T}}
\DefMacro{java_old}{\ensuremath{M_{\UseMacro{source};old}}}
\DefMacro{cs_old}{\ensuremath{M_{\UseMacro{target};old}}}
\DefMacro{java_new}{\ensuremath{M_{\UseMacro{source};new}}}
\DefMacro{cs_new}{\ensuremath{M_{\UseMacro{target};new}}}
\DefMacro{code_old}{\ensuremath{M_{old}}}
\DefMacro{code_new}{\ensuremath{M_{new}}}
\DefMacro{code_java}{\ensuremath{M_{\UseMacro{source}}}}
\DefMacro{code_cs}{\ensuremath{M_{\UseMacro{target}}}}
\DefMacro{java_edits}{\ensuremath{E_{\UseMacro{source}}}}
\DefMacro{cs_edits}{\ensuremath{E_{\UseMacro{target}}}}

\DefMacro{JavaExampleMethod}{\CodeIn{doc}\-\CodeIn{With}\-\CodeIn{Invalid}\-\CodeIn{Mapp}\-\CodeIn{ing02}\xspace}
\DefMacro{CsExampleMethod}{\CodeIn{Doc}\-\CodeIn{With}\-\CodeIn{Invalid}\-\CodeIn{Mapp}\-\CodeIn{ing02}\xspace}

\newcommand{\Java}{Java\xspace}
\newcommand{\Cs}{C\#\xspace}

\newcommand{\TrainRatio}{70\%}
\newcommand{\ValRatio}{10\%}

\newcommand{\TokenDiff}{4}
\newcommand{\EditAct}{3}
\newcommand{\ProjectsNum}{8}
\newcommand{\Datasize}{6,613}

\newcommand{\BestCodeBLEU}{96}
\newcommand{\CodeBLEUImproved}{25\%}
\newcommand{\BestImproved}{77\%}

\newcommand{\JavaInsertLineNumber}{7}
\newcommand{\JavaFirstReplaceLineNumber}{5}
\newcommand{\JavaSecondReplaceLineNumber}{6}

\newcommand{\tokenJaccardThreshold}{0.4}
\newcommand{\lineJaccardThreshold}{0.5}

\DefMacro{TCap-edits-map}{The mappings between \concise edit sequence and \unambig edit sequence.  \label{tab:edit-sequence} \vspace{-0pt}}
\DefMacro{TCap-java2cs-translation-dataset}{Statistics of the Java to Csharp translation dataset.}
\DefMacro{TCap-significance-test}{The results with the same suffixes (e.g., $\beta$) are NOT statistically significantly different.}
\DefMacro{Models-Desciptions}{*-Translation models are machine translation models, given the new Java method, they are trained to translate to new C\# code; CoditT5-Translation generates edits that convert Java to C\#; CodeT5-Update generates new C\# based on new Java code and old C\# code; CoditT5 generates C\# edits together with new C\# based on old C\#, Java edits and new Java}
\DefMacro{TCap-java2cs-translation-models-results}{Results on the \UseMacro{java2cs} dataset. \UseMacro{TCap-significance-test} \label{tab:java2cstranslation-results} \vspace{-0pt}}
\DefMacro{TCap-java2cs-cp-translation-models-results}{Results on the cross-project split using \UseMacro{java2cs} dataset. \UseMacro{TCap-significance-test} \label{tab:java2cstranslation-cp-results} \vspace{-0pt}}
\DefMacro{TCap-cs2java-cp-translation-models-results}{Results on the cross-project split using \UseMacro{cs2java} dataset. \UseMacro{TCap-significance-test} \label{tab:java2cstranslation-cp-results} \vspace{-0pt}}

\DefMacro{TCap-cs2java-translation-models-results}{Results on the \UseMacro{cs2java} dataset. \UseMacro{TCap-significance-test} \label{tab:cs2javatranslation-results} \vspace{-0pt}}
\DefMacro{TCap-java2cs-ag-translation-models-results}{Results of the Java to Csharp translation task trained with augmented data.}
\DefMacro{TCap-java2cs-new-test-translation-models-results}{Results of the Java to C\# translation on jts project (unseen)}

\DefMacro{TCap-clean-test-java2cs-translation-models-results}{Results of models on the \textbf{Clean Test} set. \UseMacro{Models-Desciptions}}
\DefMacro{TCap-real-world-dataset-stats}{Open-source projects used in our dataset and number of examples from each project.\label{tab:real-world-dataset-stats}\vspace{-3pt}}
\DefMacro{TCap-dataset-edits-stats}{Statistics of our dataset. Number of examples of training, validation and test data; average number of tokens in the old version of method and new version of method; average number of edits for the code change; average number of added and deleted tokens.\label{tab:dataset-edits-stats}\vspace{-3pt}}

\DefMacro{TVspace-real-world-dataset-stats}{-15pt}
\DefMacro{TVspace-dataset-edits-stats}{-5pt}
\DefMacro{TVspace-java2cs-translation-models-results}{-5pt}
\DefMacro{TVspace-cs2java-translation-models-results}{-5pt}
\DefMacro{TVspace-edits-map}{-5pt}

\DefMacro{THead-train}{\Train\xspace}
\DefMacro{THead-valid}{\Val\xspace}
\DefMacro{THead-test}{\Test\xspace}
\DefMacro{THead-augment}{Extra\xspace}
\DefMacro{THead-cleaned-test}{Clean Test\xspace}
\DefMacro{THead-semi}{Synthetic \xspace}
\DefMacro{THead-test-clean}{Clean Test\xspace}
\DefMacro{THead-all}{All\xspace}

\DefMacro{THead-Acc-top-1}{xMatch\xspace}
\DefMacro{THead-BLEU}{BLEU-4\xspace}
\DefMacro{THead-CodeBLEU}{CodeBLEU\xspace}
\DefMacro{THead-SARI}{SARI\xspace}
\DefMacro{THead-GLEU}{GLEU\xspace}
\DefMacro{THead-ngram_match_score}{Corpus-BLEU}

\DefMacro{THead-models}{Models\xspace}
\DefMacro{THead-weighted_ngram_match_score}{W. N-gram}
\DefMacro{THead-syntax_match_score}{Syntax}
\DefMacro{THead-dataflow_match_score}{Dfg}

\DefMacro{THead-translation-CodeT5}{\translationModel\xspace}
\DefMacro{THead-translation-CodeT5-ag}{CodeT5-Translation-\emph{extra}\xspace}
\DefMacro{THead-update-CodeT5}{\UseMacro{CodeT5Up}\xspace}
\DefMacro{THead-CopyEdits}{\CopyEdits\xspace}
\DefMacro{THead-metaEdits}{\DeltaTool(MetaEdits)}
\DefMacro{THead-edit-translation}{\DeltaTool(EditsTranslation)}
\DefMacro{THead-editModel}{Edit Model\xspace}
\DefMacro{THead-update-CodeT5-ag}{CodeT5-Update-\emph{extra}\xspace}
\DefMacro{THead-hybrid}{Hybrid}
\DefMacro{THead-chatgpt3.5-few-shot}{ChatGPT-few-shot\xspace}
\DefMacro{THead-chatgpt3.5-zero-shot}{ChatGPT-zero-shot\xspace}

\DefMacro{THead-translation-CoditT5}{CoditT5-Translation\xspace}
\DefMacro{THead-CoditT5}{CoditT5\xspace}
\DefMacro{THead-CoditT5-ag}{CoditT5-\emph{extra}\xspace}
\DefMacro{THead-copy}{Copy\xspace}
\DefMacro{THead-codex}{Codex-few-shot\xspace}

\DefMacro{THead-count}{Count\xspace}
\DefMacro{THead-java-project}{Java Project\xspace}
\DefMacro{THead-cs-project}{C\# Project\xspace}
\DefMacro{THead-train-commit}{\Train commit dates\xspace}
\DefMacro{THead-valid-commit}{\Val commit dates\xspace}
\DefMacro{THead-test-commit}{\Test commit dates\xspace}
\DefMacro{THead-cs-edit-distance-mean}{Edit distance for C\#}
\DefMacro{THead-java-edit-distance-mean}{Edit distance for Java}

\DefMacro{THead-mean-old-method-size}{Avg. len(\UseMacro{code_old})}
\DefMacro{THead-mean-new-method-size}{Avg. len(\UseMacro{code_new})}
\DefMacro{THead-mean-edit-actions}{Avg. \# edits}
\DefMacro{THead-mean-edit-lines}{Avg. \# edited lines}
\DefMacro{THead-mean-edit-tokens}{Avg. \# tks}
\DefMacro{THead-mean-add-tokens}{Avg. \# add. tks}
\DefMacro{THead-mean-del-tokens}{Avg. \# del. tks}

\DefMacro{ds-test-count}{1,599}
\DefMacro{ds-test-cs-edit-distance-mean}{17.88}
\DefMacro{ds-test-java-edit-distance-mean}{16.40}
\DefMacro{ds-train-count}{4,391}
\DefMacro{ds-train-cs-edit-distance-mean}{28.59}
\DefMacro{ds-train-java-edit-distance-mean}{27.59}
\DefMacro{ds-valid-count}{623}
\DefMacro{ds-valid-cs-edit-distance-mean}{27.38}
\DefMacro{ds-valid-java-edit-distance-mean}{25.73}

\DefMacro{res-java2cs-copy-Acc-top-1}{0.00}
\DefMacro{res-java2cs-copy-BLEU}{83.11}
\DefMacro{res-java2cs-copy-COPY-PCT}{100.00}
\DefMacro{res-java2cs-copy-CodeBLEU}{90.42}
\DefMacro{res-java2cs-copy-GLEU}{74.58}
\DefMacro{res-java2cs-copy-SARI}{30.68}
\DefMacro{res-java2cs-copy-dataflow_match_score}{84.95}
\DefMacro{res-java2cs-copy-ngram_match_score}{92.12}
\DefMacro{res-java2cs-copy-syntax_match_score}{92.45}
\DefMacro{res-java2cs-copy-weighted_ngram_match_score}{92.17}
\DefMacro{res-java2cs-CopyEdits-Acc-top-1}{38.21}
\DefMacro{res-java2cs-CopyEdits-BLEU}{90.29}
\DefMacro{res-java2cs-CopyEdits-COPY-PCT}{5.57}
\DefMacro{res-java2cs-CopyEdits-CodeBLEU}{91.34}
\DefMacro{res-java2cs-CopyEdits-GLEU}{87.93}
\DefMacro{res-java2cs-CopyEdits-SARI}{76.92}
\DefMacro{res-java2cs-CopyEdits-dataflow_match_score}{85.13}
\DefMacro{res-java2cs-CopyEdits-ngram_match_score}{92.45}
\DefMacro{res-java2cs-CopyEdits-syntax_match_score}{92.48}
\DefMacro{res-java2cs-CopyEdits-weighted_ngram_match_score}{95.29}
\DefMacro{res-java2cs-translation-CodeT5-Acc-top-1}{38.02}
\DefMacro{res-java2cs-translation-CodeT5-BLEU}{87.45}
\DefMacro{res-java2cs-translation-CodeT5-COPY-PCT}{0.00}
\DefMacro{res-java2cs-translation-CodeT5-CodeBLEU}{77.15}
\DefMacro{res-java2cs-translation-CodeT5-GLEU}{85.59}
\DefMacro{res-java2cs-translation-CodeT5-SARI}{83.77}
\DefMacro{res-java2cs-translation-CodeT5-dataflow_match_score}{68.24}
\DefMacro{res-java2cs-translation-CodeT5-ngram_match_score}{76.61}
\DefMacro{res-java2cs-translation-CodeT5-syntax_match_score}{85.53}
\DefMacro{res-java2cs-translation-CodeT5-weighted_ngram_match_score}{78.21}
\DefMacro{res-java2cs-codex-Acc-top-1}{48.84}
\DefMacro{res-java2cs-codex-BLEU}{80.71}
\DefMacro{res-java2cs-codex-COPY-PCT}{4.13}
\DefMacro{res-java2cs-codex-CodeBLEU}{59.63}
\DefMacro{res-java2cs-codex-GLEU}{79.74}
\DefMacro{res-java2cs-codex-SARI}{72.80}
\DefMacro{res-java2cs-codex-dataflow_match_score}{49.78}
\DefMacro{res-java2cs-codex-ngram_match_score}{51.91}
\DefMacro{res-java2cs-codex-syntax_match_score}{76.71}
\DefMacro{res-java2cs-codex-weighted_ngram_match_score}{60.12}
\DefMacro{res-java2cs-CoditT5-Acc-top-1}{60.29}
\DefMacro{res-java2cs-CoditT5-BLEU}{89.84}
\DefMacro{res-java2cs-CoditT5-COPY-PCT}{0.00}
\DefMacro{res-java2cs-CoditT5-CodeBLEU}{75.20}
\DefMacro{res-java2cs-CoditT5-GLEU}{89.29}
\DefMacro{res-java2cs-CoditT5-SARI}{80.99}
\DefMacro{res-java2cs-CoditT5-dataflow_match_score}{67.76}
\DefMacro{res-java2cs-CoditT5-ngram_match_score}{72.70}
\DefMacro{res-java2cs-CoditT5-syntax_match_score}{84.99}
\DefMacro{res-java2cs-CoditT5-weighted_ngram_match_score}{75.35}
\DefMacro{res-java2cs-update-CodeT5-Acc-top-1}{60.41}
\DefMacro{res-java2cs-update-CodeT5-BLEU}{90.00}
\DefMacro{res-java2cs-update-CodeT5-COPY-PCT}{4.50}
\DefMacro{res-java2cs-update-CodeT5-CodeBLEU}{76.63}
\DefMacro{res-java2cs-update-CodeT5-GLEU}{88.72}
\DefMacro{res-java2cs-update-CodeT5-SARI}{80.11}
\DefMacro{res-java2cs-update-CodeT5-dataflow_match_score}{68.92}
\DefMacro{res-java2cs-update-CodeT5-ngram_match_score}{75.19}
\DefMacro{res-java2cs-update-CodeT5-syntax_match_score}{85.74}
\DefMacro{res-java2cs-update-CodeT5-weighted_ngram_match_score}{76.66}
\DefMacro{res-java2cs-chatgpt3.5-few-shot-Acc-top-1}{63.16}
\DefMacro{res-java2cs-chatgpt3.5-few-shot-BLEU}{92.23}
\DefMacro{res-java2cs-chatgpt3.5-few-shot-COPY-PCT}{2.00}
\DefMacro{res-java2cs-chatgpt3.5-few-shot-CodeBLEU}{80.57}
\DefMacro{res-java2cs-chatgpt3.5-few-shot-GLEU}{91.13}
\DefMacro{res-java2cs-chatgpt3.5-few-shot-SARI}{82.06}
\DefMacro{res-java2cs-chatgpt3.5-few-shot-dataflow_match_score}{73.20}
\DefMacro{res-java2cs-chatgpt3.5-few-shot-ngram_match_score}{79.47}
\DefMacro{res-java2cs-chatgpt3.5-few-shot-syntax_match_score}{88.73}
\DefMacro{res-java2cs-chatgpt3.5-few-shot-weighted_ngram_match_score}{80.89}
\DefMacro{res-java2cs-chatgpt3.5-zero-shot-Acc-top-1}{29.52}
\DefMacro{res-java2cs-chatgpt3.5-zero-shot-BLEU}{85.60}
\DefMacro{res-java2cs-chatgpt3.5-zero-shot-COPY-PCT}{1.06}
\DefMacro{res-java2cs-chatgpt3.5-zero-shot-CodeBLEU}{73.00}
\DefMacro{res-java2cs-chatgpt3.5-zero-shot-GLEU}{84.74}
\DefMacro{res-java2cs-chatgpt3.5-zero-shot-SARI}{68.44}
\DefMacro{res-java2cs-chatgpt3.5-zero-shot-dataflow_match_score}{62.53}
\DefMacro{res-java2cs-chatgpt3.5-zero-shot-ngram_match_score}{76.81}
\DefMacro{res-java2cs-chatgpt3.5-zero-shot-syntax_match_score}{74.74}
\DefMacro{res-java2cs-chatgpt3.5-zero-shot-weighted_ngram_match_score}{77.91}
\DefMacro{res-java2cs-metaEdits-Acc-top-1}{63.48}
\DefMacro{res-java2cs-metaEdits-BLEU}{94.55}
\DefMacro{res-java2cs-metaEdits-COPY-PCT}{0.50}
\DefMacro{res-java2cs-metaEdits-CodeBLEU}{94.78}
\DefMacro{res-java2cs-metaEdits-GLEU}{93.20}
\DefMacro{res-java2cs-metaEdits-SARI}{85.63}
\DefMacro{res-java2cs-metaEdits-dataflow_match_score}{91.47}
\DefMacro{res-java2cs-metaEdits-ngram_match_score}{95.50}
\DefMacro{res-java2cs-metaEdits-syntax_match_score}{95.15}
\DefMacro{res-java2cs-metaEdits-weighted_ngram_match_score}{97.01}
\DefMacro{res-java2cs-edit-translation-Acc-top-1}{67.23}
\DefMacro{res-java2cs-edit-translation-BLEU}{95.44}
\DefMacro{res-java2cs-edit-translation-COPY-PCT}{0.31}
\DefMacro{res-java2cs-edit-translation-CodeBLEU}{96.02}
\DefMacro{res-java2cs-edit-translation-GLEU}{94.21}
\DefMacro{res-java2cs-edit-translation-SARI}{87.23}
\DefMacro{res-java2cs-edit-translation-dataflow_match_score}{93.34}
\DefMacro{res-java2cs-edit-translation-ngram_match_score}{96.56}
\DefMacro{res-java2cs-edit-translation-syntax_match_score}{96.32}
\DefMacro{res-java2cs-edit-translation-weighted_ngram_match_score}{97.85}
\DefMacro{res-java2cs-hybrid-Acc-top-1}{71.79}
\DefMacro{res-java2cs-hybrid-BLEU}{96.12}
\DefMacro{res-java2cs-hybrid-COPY-PCT}{1.94}
\DefMacro{res-java2cs-hybrid-CodeBLEU}{96.09}
\DefMacro{res-java2cs-hybrid-GLEU}{95.07}
\DefMacro{res-java2cs-hybrid-SARI}{87.08}
\DefMacro{res-java2cs-hybrid-dataflow_match_score}{93.38}
\DefMacro{res-java2cs-hybrid-ngram_match_score}{96.62}
\DefMacro{res-java2cs-hybrid-syntax_match_score}{96.64}
\DefMacro{res-java2cs-hybrid-weighted_ngram_match_score}{97.74}

\DefMacro{res-java2cs-ag-translation-CodeT5-Acc-top-1}{35.17}
\DefMacro{res-java2cs-ag-translation-CodeT5-BLEU}{69.99}
\DefMacro{res-java2cs-ag-translation-CodeT5-COPY-PCT}{0.00}
\DefMacro{res-java2cs-ag-translation-CodeT5-CodeBLEU}{65.45}
\DefMacro{res-java2cs-ag-update-CodeT5-Acc-top-1}{41.72}
\DefMacro{res-java2cs-ag-update-CodeT5-BLEU}{81.46}
\DefMacro{res-java2cs-ag-update-CodeT5-COPY-PCT}{0.00}
\DefMacro{res-java2cs-ag-update-CodeT5-CodeBLEU}{72.55}

\DefMacro{res-clean-test-java2cs-translation-CodeT5-Acc-top-1}{55.62}
\DefMacro{res-clean-test-java2cs-translation-CodeT5-BLEU}{91.68}
\DefMacro{res-clean-test-java2cs-translation-CodeT5-COPY-PCT}{0.00}
\DefMacro{res-clean-test-java2cs-translation-CodeT5-CodeBLEU}{84.34}
\DefMacro{res-clean-test-java2cs-translation-CodeT5-ag-Acc-top-1}{58.99}
\DefMacro{res-clean-test-java2cs-translation-CodeT5-ag-BLEU}{91.46}
\DefMacro{res-clean-test-java2cs-translation-CodeT5-ag-COPY-PCT}{0.00}
\DefMacro{res-clean-test-java2cs-translation-CodeT5-ag-CodeBLEU}{84.66}
\DefMacro{res-clean-test-java2cs-translation-CoditT5-Acc-top-1}{55.62}
\DefMacro{res-clean-test-java2cs-translation-CoditT5-BLEU}{90.24}
\DefMacro{res-clean-test-java2cs-translation-CoditT5-COPY-PCT}{0.00}
\DefMacro{res-clean-test-java2cs-translation-CoditT5-CodeBLEU}{82.23}
\DefMacro{res-clean-test-java2cs-update-CodeT5-Acc-top-1}{64.61}
\DefMacro{res-clean-test-java2cs-update-CodeT5-BLEU}{94.62}
\DefMacro{res-clean-test-java2cs-update-CodeT5-COPY-PCT}{8.99}
\DefMacro{res-clean-test-java2cs-update-CodeT5-CodeBLEU}{90.59}
\DefMacro{res-clean-test-java2cs-update-CodeT5-ag-Acc-top-1}{71.91}
\DefMacro{res-clean-test-java2cs-update-CodeT5-ag-BLEU}{97.30}
\DefMacro{res-clean-test-java2cs-update-CodeT5-ag-COPY-PCT}{10.11}
\DefMacro{res-clean-test-java2cs-update-CodeT5-ag-CodeBLEU}{94.07}
\DefMacro{res-clean-test-java2cs-CoditT5-Acc-top-1}{67.98}
\DefMacro{res-clean-test-java2cs-CoditT5-BLEU}{95.25}
\DefMacro{res-clean-test-java2cs-CoditT5-COPY-PCT}{10.67}
\DefMacro{res-clean-test-java2cs-CoditT5-CodeBLEU}{91.30}
\DefMacro{res-clean-test-java2cs-CoditT5-ag-Acc-top-1}{66.85}
\DefMacro{res-clean-test-java2cs-CoditT5-ag-BLEU}{95.11}
\DefMacro{res-clean-test-java2cs-CoditT5-ag-COPY-PCT}{11.80}
\DefMacro{res-clean-test-java2cs-CoditT5-ag-CodeBLEU}{91.37}
\DefMacro{res-clean-test-java2cs-copy-Acc-top-1}{0.00}
\DefMacro{res-clean-test-java2cs-copy-BLEU}{80.84}
\DefMacro{res-clean-test-java2cs-copy-COPY-PCT}{100.00}
\DefMacro{res-clean-test-java2cs-copy-CodeBLEU}{83.02}

\DefMacro{antlr-antlr4-count}{ 12}
\DefMacro{antlr-antlr4-cs-test-commit}{2017-01-06 19:40:21}
\DefMacro{antlr-antlr4-cs-train-commit}{2013-08-04 22:01:15}
\DefMacro{antlr-antlr4-cs-valid-commit}{2014-03-25 14:48:19}
\DefMacro{antlr-antlr4-java-test-commit}{2016-12-08 11:54:07}
\DefMacro{antlr-antlr4-java-train-commit}{2013-06-08 21:09:33}
\DefMacro{antlr-antlr4-java-valid-commit}{2014-02-18 12:23:42}
\DefMacro{apache-lucene-count}{ 40}
\DefMacro{apache-lucene-cs-test-commit}{2012-09-22 23:04:38}
\DefMacro{apache-lucene-cs-train-commit}{2006-08-17 13:49:26}
\DefMacro{apache-lucene-cs-valid-commit}{2012-09-22 17:48:59}
\DefMacro{apache-lucene-java-test-commit}{2012-09-05 15:37:11}
\DefMacro{apache-lucene-java-train-commit}{2006-05-23 15:01:11}
\DefMacro{apache-lucene-java-valid-commit}{2012-09-15 14:49:24}
\DefMacro{apache-poi-count}{ 5}
\DefMacro{apache-poi-cs-test-commit}{2013-01-12 15:51:00}
\DefMacro{apache-poi-cs-train-commit}{2011-10-04 08:42:22}
\DefMacro{apache-poi-cs-valid-commit}{2013-01-10 21:54:31}
\DefMacro{apache-poi-java-test-commit}{2012-10-25 15:13:53}
\DefMacro{apache-poi-java-train-commit}{2011-07-19 12:19:48}
\DefMacro{apache-poi-java-valid-commit}{2012-10-25 15:13:53}
\DefMacro{eclipse-jgit-count}{ 808}
\DefMacro{eclipse-jgit-cs-test-commit}{2012-07-03 19:30:14}
\DefMacro{eclipse-jgit-cs-train-commit}{2010-10-14 15:53:43}
\DefMacro{eclipse-jgit-cs-valid-commit}{2012-03-08 11:00:24}
\DefMacro{eclipse-jgit-java-test-commit}{2012-05-04 14:37:40}
\DefMacro{eclipse-jgit-java-train-commit}{2010-10-11 16:20:00}
\DefMacro{eclipse-jgit-java-valid-commit}{2011-12-16 17:05:47}
\DefMacro{formicary-fpml-toolkit-java-count}{ 20}
\DefMacro{formicary-fpml-toolkit-java-cs-test-commit}{2008-04-04 20:01:19}
\DefMacro{formicary-fpml-toolkit-java-cs-train-commit}{2007-07-29 08:07:31}
\DefMacro{formicary-fpml-toolkit-java-cs-valid-commit}{2007-10-29 22:38:05}
\DefMacro{formicary-fpml-toolkit-java-java-test-commit}{2008-04-04 19:57:59}
\DefMacro{formicary-fpml-toolkit-java-java-train-commit}{2007-07-29 08:06:22}
\DefMacro{formicary-fpml-toolkit-java-java-valid-commit}{2007-10-29 22:32:02}
\DefMacro{itext-itext7-count}{ 5,121}
\DefMacro{itext-itext7-cs-test-commit}{2021-07-26 17:33:15}
\DefMacro{itext-itext7-cs-train-commit}{2016-04-26 20:59:42}
\DefMacro{itext-itext7-cs-valid-commit}{2020-06-12 10:03:17}
\DefMacro{itext-itext7-java-test-commit}{2021-07-26 17:26:53}
\DefMacro{itext-itext7-java-train-commit}{2016-04-22 14:00:15}
\DefMacro{itext-itext7-java-valid-commit}{2020-06-12 09:42:37}
\DefMacro{quartz-scheduler-quartz-count}{ 17}
\DefMacro{quartz-scheduler-quartz-cs-test-commit}{2015-01-11 18:59:18}
\DefMacro{quartz-scheduler-quartz-cs-train-commit}{2012-06-23 23:00:07}
\DefMacro{quartz-scheduler-quartz-cs-valid-commit}{2013-11-02 22:42:16}
\DefMacro{quartz-scheduler-quartz-java-test-commit}{2014-11-11 02:24:50}
\DefMacro{quartz-scheduler-quartz-java-train-commit}{2012-05-28 15:36:36}
\DefMacro{quartz-scheduler-quartz-java-valid-commit}{2013-09-16 15:19:49}
\DefMacro{terabyte-jgit-count}{ 590}
\DefMacro{terabyte-jgit-cs-test-commit}{2011-03-08 17:03:09}
\DefMacro{terabyte-jgit-cs-train-commit}{2010-10-14 15:53:43}
\DefMacro{terabyte-jgit-cs-valid-commit}{2011-02-04 11:29:58}
\DefMacro{terabyte-jgit-java-test-commit}{2011-03-07 18:23:39}
\DefMacro{terabyte-jgit-java-train-commit}{2010-10-11 16:20:00}
\DefMacro{terabyte-jgit-java-valid-commit}{2011-02-02 17:16:32}

\DefMacro{ds-total-count}{ 4,736}
\DefMacro{ds-cs-edit-distance-mean}{ 12.82}
\DefMacro{ds-java-edit-distance-mean}{ 11.65}
\DefMacro{ds-antlr-antlr4-count}{ 17}
\DefMacro{ds-apache-lucene-count}{ 1,331}
\DefMacro{ds-apache-poi-count}{ 368}
\DefMacro{ds-eclipse-jgit-count}{ 395}
\DefMacro{ds-formicary-fpml-toolkit-java-count}{ 5}
\DefMacro{ds-itext-itext7-count}{ 2,296}
\DefMacro{ds-quartz-scheduler-quartz-count}{ 92}
\DefMacro{ds-terabyte-jgit-count}{ 232}

\DefMacro{res-java2cs-new-test-copy-Acc-top-1}{0.00}
\DefMacro{res-java2cs-new-test-copy-BLEU}{80.73}
\DefMacro{res-java2cs-new-test-copy-COPY-PCT}{100.00}
\DefMacro{res-java2cs-new-test-copy-CodeBLEU}{85.41}
\DefMacro{res-java2cs-new-test-copy-dataflow_match_score}{83.37}
\DefMacro{res-java2cs-new-test-copy-ngram_match_score}{85.16}
\DefMacro{res-java2cs-new-test-copy-syntax_match_score}{85.30}
\DefMacro{res-java2cs-new-test-copy-weighted_ngram_match_score}{87.79}
\DefMacro{res-java2cs-new-test-CopyEdits-Acc-top-1}{27.78}
\DefMacro{res-java2cs-new-test-CopyEdits-BLEU}{87.73}
\DefMacro{res-java2cs-new-test-CopyEdits-COPY-PCT}{13.89}
\DefMacro{res-java2cs-new-test-CopyEdits-CodeBLEU}{86.28}
\DefMacro{res-java2cs-new-test-CopyEdits-dataflow_match_score}{79.07}
\DefMacro{res-java2cs-new-test-CopyEdits-ngram_match_score}{86.76}
\DefMacro{res-java2cs-new-test-CopyEdits-syntax_match_score}{87.38}
\DefMacro{res-java2cs-new-test-CopyEdits-weighted_ngram_match_score}{91.92}
\DefMacro{res-java2cs-new-test-translation-CodeT5-Acc-top-1}{2.78}
\DefMacro{res-java2cs-new-test-translation-CodeT5-BLEU}{75.68}
\DefMacro{res-java2cs-new-test-translation-CodeT5-COPY-PCT}{0.00}
\DefMacro{res-java2cs-new-test-translation-CodeT5-CodeBLEU}{75.26}
\DefMacro{res-java2cs-new-test-translation-CodeT5-dataflow_match_score}{74.64}
\DefMacro{res-java2cs-new-test-translation-CodeT5-ngram_match_score}{74.51}
\DefMacro{res-java2cs-new-test-translation-CodeT5-syntax_match_score}{73.74}
\DefMacro{res-java2cs-new-test-translation-CodeT5-weighted_ngram_match_score}{78.15}
\DefMacro{res-java2cs-new-test-CoditT5-Acc-top-1}{44.44}
\DefMacro{res-java2cs-new-test-CoditT5-BLEU}{90.97}
\DefMacro{res-java2cs-new-test-CoditT5-COPY-PCT}{2.78}
\DefMacro{res-java2cs-new-test-CoditT5-CodeBLEU}{87.03}
\DefMacro{res-java2cs-new-test-CoditT5-dataflow_match_score}{84.75}
\DefMacro{res-java2cs-new-test-CoditT5-ngram_match_score}{87.35}
\DefMacro{res-java2cs-new-test-CoditT5-syntax_match_score}{88.46}
\DefMacro{res-java2cs-new-test-CoditT5-weighted_ngram_match_score}{87.55}
\DefMacro{res-java2cs-new-test-update-CodeT5-Acc-top-1}{38.89}
\DefMacro{res-java2cs-new-test-update-CodeT5-BLEU}{90.62}
\DefMacro{res-java2cs-new-test-update-CodeT5-COPY-PCT}{0.00}
\DefMacro{res-java2cs-new-test-update-CodeT5-CodeBLEU}{87.78}
\DefMacro{res-java2cs-new-test-update-CodeT5-dataflow_match_score}{84.81}
\DefMacro{res-java2cs-new-test-update-CodeT5-ngram_match_score}{87.37}
\DefMacro{res-java2cs-new-test-update-CodeT5-syntax_match_score}{89.69}
\DefMacro{res-java2cs-new-test-update-CodeT5-weighted_ngram_match_score}{89.24}
\DefMacro{res-java2cs-new-test-metaEdits-Acc-top-1}{41.67}
\DefMacro{res-java2cs-new-test-metaEdits-BLEU}{89.04}
\DefMacro{res-java2cs-new-test-metaEdits-COPY-PCT}{5.56}
\DefMacro{res-java2cs-new-test-metaEdits-CodeBLEU}{86.02}
\DefMacro{res-java2cs-new-test-metaEdits-dataflow_match_score}{77.33}
\DefMacro{res-java2cs-new-test-metaEdits-ngram_match_score}{87.96}
\DefMacro{res-java2cs-new-test-metaEdits-syntax_match_score}{86.23}
\DefMacro{res-java2cs-new-test-metaEdits-weighted_ngram_match_score}{92.57}
\DefMacro{res-java2cs-new-test-edit-translation-Acc-top-1}{37.50}
\DefMacro{res-java2cs-new-test-edit-translation-BLEU}{89.76}
\DefMacro{res-java2cs-new-test-edit-translation-COPY-PCT}{5.56}
\DefMacro{res-java2cs-new-test-edit-translation-CodeBLEU}{87.58}
\DefMacro{res-java2cs-new-test-edit-translation-dataflow_match_score}{80.80}
\DefMacro{res-java2cs-new-test-edit-translation-ngram_match_score}{88.03}
\DefMacro{res-java2cs-new-test-edit-translation-syntax_match_score}{88.51}
\DefMacro{res-java2cs-new-test-edit-translation-weighted_ngram_match_score}{92.98}

\DefMacro{res-cs2java-copy-Acc-top-1}{0.00}
\DefMacro{res-cs2java-copy-BLEU}{83.06}
\DefMacro{res-cs2java-copy-COPY-PCT}{100.00}
\DefMacro{res-cs2java-copy-CodeBLEU}{89.82}
\DefMacro{res-cs2java-copy-GLEU}{74.55}
\DefMacro{res-cs2java-copy-SARI}{30.66}
\DefMacro{res-cs2java-copy-dataflow_match_score}{82.08}
\DefMacro{res-cs2java-copy-ngram_match_score}{92.29}
\DefMacro{res-cs2java-copy-syntax_match_score}{92.59}
\DefMacro{res-cs2java-copy-weighted_ngram_match_score}{92.31}
\DefMacro{res-cs2java-CopyEdits-Acc-top-1}{38.15}
\DefMacro{res-cs2java-CopyEdits-BLEU}{89.36}
\DefMacro{res-cs2java-CopyEdits-COPY-PCT}{6.07}
\DefMacro{res-cs2java-CopyEdits-CodeBLEU}{90.31}
\DefMacro{res-cs2java-CopyEdits-GLEU}{87.02}
\DefMacro{res-cs2java-CopyEdits-SARI}{75.86}
\DefMacro{res-cs2java-CopyEdits-dataflow_match_score}{82.21}
\DefMacro{res-cs2java-CopyEdits-ngram_match_score}{92.16}
\DefMacro{res-cs2java-CopyEdits-syntax_match_score}{91.98}
\DefMacro{res-cs2java-CopyEdits-weighted_ngram_match_score}{94.91}
\DefMacro{res-cs2java-translation-CodeT5-Acc-top-1}{40.21}
\DefMacro{res-cs2java-translation-CodeT5-BLEU}{89.10}
\DefMacro{res-cs2java-translation-CodeT5-COPY-PCT}{0.00}
\DefMacro{res-cs2java-translation-CodeT5-CodeBLEU}{77.99}
\DefMacro{res-cs2java-translation-CodeT5-GLEU}{87.21}
\DefMacro{res-cs2java-translation-CodeT5-SARI}{83.99}
\DefMacro{res-cs2java-translation-CodeT5-dataflow_match_score}{68.07}
\DefMacro{res-cs2java-translation-CodeT5-ngram_match_score}{78.00}
\DefMacro{res-cs2java-translation-CodeT5-syntax_match_score}{86.55}
\DefMacro{res-cs2java-translation-CodeT5-weighted_ngram_match_score}{79.34}
\DefMacro{res-cs2java-codex-Acc-top-1}{55.53}
\DefMacro{res-cs2java-codex-BLEU}{82.54}
\DefMacro{res-cs2java-codex-COPY-PCT}{1.63}
\DefMacro{res-cs2java-codex-CodeBLEU}{60.35}
\DefMacro{res-cs2java-codex-GLEU}{82.13}
\DefMacro{res-cs2java-codex-SARI}{76.23}
\DefMacro{res-cs2java-codex-dataflow_match_score}{51.01}
\DefMacro{res-cs2java-codex-ngram_match_score}{52.87}
\DefMacro{res-cs2java-codex-syntax_match_score}{76.78}
\DefMacro{res-cs2java-codex-weighted_ngram_match_score}{60.73}
\DefMacro{res-cs2java-CoditT5-Acc-top-1}{60.98}
\DefMacro{res-cs2java-CoditT5-BLEU}{90.88}
\DefMacro{res-cs2java-CoditT5-COPY-PCT}{0.00}
\DefMacro{res-cs2java-CoditT5-CodeBLEU}{75.87}
\DefMacro{res-cs2java-CoditT5-GLEU}{90.15}
\DefMacro{res-cs2java-CoditT5-SARI}{81.41}
\DefMacro{res-cs2java-CoditT5-dataflow_match_score}{67.66}
\DefMacro{res-cs2java-CoditT5-ngram_match_score}{74.03}
\DefMacro{res-cs2java-CoditT5-syntax_match_score}{85.50}
\DefMacro{res-cs2java-CoditT5-weighted_ngram_match_score}{76.27}
\DefMacro{res-cs2java-update-CodeT5-Acc-top-1}{55.97}
\DefMacro{res-cs2java-update-CodeT5-BLEU}{90.62}
\DefMacro{res-cs2java-update-CodeT5-COPY-PCT}{3.19}
\DefMacro{res-cs2java-update-CodeT5-CodeBLEU}{76.38}
\DefMacro{res-cs2java-update-CodeT5-GLEU}{89.72}
\DefMacro{res-cs2java-update-CodeT5-SARI}{79.65}
\DefMacro{res-cs2java-update-CodeT5-dataflow_match_score}{67.86}
\DefMacro{res-cs2java-update-CodeT5-ngram_match_score}{75.20}
\DefMacro{res-cs2java-update-CodeT5-syntax_match_score}{85.86}
\DefMacro{res-cs2java-update-CodeT5-weighted_ngram_match_score}{76.60}
\DefMacro{res-cs2java-chatgpt3.5-few-shot-Acc-top-1}{70.92}
\DefMacro{res-cs2java-chatgpt3.5-few-shot-BLEU}{94.43}
\DefMacro{res-cs2java-chatgpt3.5-few-shot-COPY-PCT}{2.75}
\DefMacro{res-cs2java-chatgpt3.5-few-shot-CodeBLEU}{82.77}
\DefMacro{res-cs2java-chatgpt3.5-few-shot-GLEU}{93.80}
\DefMacro{res-cs2java-chatgpt3.5-few-shot-SARI}{83.63}
\DefMacro{res-cs2java-chatgpt3.5-few-shot-dataflow_match_score}{74.89}
\DefMacro{res-cs2java-chatgpt3.5-few-shot-ngram_match_score}{82.27}
\DefMacro{res-cs2java-chatgpt3.5-few-shot-syntax_match_score}{90.67}
\DefMacro{res-cs2java-chatgpt3.5-few-shot-weighted_ngram_match_score}{83.25}
\DefMacro{res-cs2java-chatgpt3.5-zero-shot-Acc-top-1}{32.52}
\DefMacro{res-cs2java-chatgpt3.5-zero-shot-BLEU}{86.95}
\DefMacro{res-cs2java-chatgpt3.5-zero-shot-COPY-PCT}{0.31}
\DefMacro{res-cs2java-chatgpt3.5-zero-shot-CodeBLEU}{76.01}
\DefMacro{res-cs2java-chatgpt3.5-zero-shot-GLEU}{86.33}
\DefMacro{res-cs2java-chatgpt3.5-zero-shot-SARI}{69.05}
\DefMacro{res-cs2java-chatgpt3.5-zero-shot-dataflow_match_score}{66.69}
\DefMacro{res-cs2java-chatgpt3.5-zero-shot-ngram_match_score}{79.97}
\DefMacro{res-cs2java-chatgpt3.5-zero-shot-syntax_match_score}{76.88}
\DefMacro{res-cs2java-chatgpt3.5-zero-shot-weighted_ngram_match_score}{80.50}
\DefMacro{res-cs2java-metaEdits-Acc-top-1}{68.61}
\DefMacro{res-cs2java-metaEdits-BLEU}{93.98}
\DefMacro{res-cs2java-metaEdits-COPY-PCT}{0.19}
\DefMacro{res-cs2java-metaEdits-CodeBLEU}{94.43}
\DefMacro{res-cs2java-metaEdits-GLEU}{92.61}
\DefMacro{res-cs2java-metaEdits-SARI}{85.74}
\DefMacro{res-cs2java-metaEdits-dataflow_match_score}{90.63}
\DefMacro{res-cs2java-metaEdits-ngram_match_score}{95.02}
\DefMacro{res-cs2java-metaEdits-syntax_match_score}{95.35}
\DefMacro{res-cs2java-metaEdits-weighted_ngram_match_score}{96.71}
\DefMacro{res-cs2java-edit-translation-Acc-top-1}{67.92}
\DefMacro{res-cs2java-edit-translation-BLEU}{95.29}
\DefMacro{res-cs2java-edit-translation-COPY-PCT}{0.50}
\DefMacro{res-cs2java-edit-translation-CodeBLEU}{94.83}
\DefMacro{res-cs2java-edit-translation-GLEU}{94.23}
\DefMacro{res-cs2java-edit-translation-SARI}{86.24}
\DefMacro{res-cs2java-edit-translation-dataflow_match_score}{89.96}
\DefMacro{res-cs2java-edit-translation-ngram_match_score}{96.10}
\DefMacro{res-cs2java-edit-translation-syntax_match_score}{95.75}
\DefMacro{res-cs2java-edit-translation-weighted_ngram_match_score}{97.51}
\DefMacro{res-cs2java-hybrid-Acc-top-1}{67.67}
\DefMacro{res-cs2java-hybrid-BLEU}{96.44}
\DefMacro{res-cs2java-hybrid-COPY-PCT}{0.00}
\DefMacro{res-cs2java-hybrid-CodeBLEU}{95.36}
\DefMacro{res-cs2java-hybrid-GLEU}{95.75}
\DefMacro{res-cs2java-hybrid-SARI}{84.46}
\DefMacro{res-cs2java-hybrid-dataflow_match_score}{90.66}
\DefMacro{res-cs2java-hybrid-ngram_match_score}{96.63}
\DefMacro{res-cs2java-hybrid-syntax_match_score}{96.50}
\DefMacro{res-cs2java-hybrid-weighted_ngram_match_score}{97.67}

\DefMacro{test-cs-mean-add-tokens}{ 11.86}
\DefMacro{test-cs-median-add-tokens}{ 2}
\DefMacro{test-cs-mean-del-tokens}{ 11.25}
\DefMacro{test-cs-median-del-tokens}{ 2}
\DefMacro{test-cs-mean-edit-actions}{ 2.47}
\DefMacro{test-cs-median-edit-actions}{ 2}
\DefMacro{test-cs-mean-edit-lines}{ 5.47}
\DefMacro{test-cs-median-edit-lines}{ 3}
\DefMacro{test-cs-mean-edit-tokens}{ 23.11}
\DefMacro{test-cs-median-edit-tokens}{ 6}
\DefMacro{test-cs-mean-new-method-size}{ 169.22}
\DefMacro{test-cs-median-new-method-size}{ 101}
\DefMacro{test-cs-mean-old-method-size}{ 168.60}
\DefMacro{test-cs-median-old-method-size}{ 100}
\DefMacro{test-java-mean-add-tokens}{ 10.90}
\DefMacro{test-java-median-add-tokens}{ 2}
\DefMacro{test-java-mean-del-tokens}{ 10.59}
\DefMacro{test-java-median-del-tokens}{ 1}
\DefMacro{test-java-mean-edit-actions}{ 2.43}
\DefMacro{test-java-median-edit-actions}{ 2}
\DefMacro{test-java-mean-edit-lines}{ 4.43}
\DefMacro{test-java-median-edit-lines}{ 2}
\DefMacro{test-java-mean-edit-tokens}{ 21.49}
\DefMacro{test-java-median-edit-tokens}{ 6}
\DefMacro{test-java-mean-new-method-size}{ 159.37}
\DefMacro{test-java-median-new-method-size}{ 94}
\DefMacro{test-java-mean-old-method-size}{ 159.06}
\DefMacro{test-java-median-old-method-size}{ 92}
\DefMacro{train-cs-mean-add-tokens}{ 20.30}
\DefMacro{train-cs-median-add-tokens}{ 4}
\DefMacro{train-cs-mean-del-tokens}{ 17.18}
\DefMacro{train-cs-median-del-tokens}{ 2}
\DefMacro{train-cs-mean-edit-actions}{ 2.73}
\DefMacro{train-cs-median-edit-actions}{ 2}
\DefMacro{train-cs-mean-edit-lines}{ 6.46}
\DefMacro{train-cs-median-edit-lines}{ 3}
\DefMacro{train-cs-mean-edit-tokens}{ 37.48}
\DefMacro{train-cs-median-edit-tokens}{ 11}
\DefMacro{train-cs-mean-new-method-size}{ 203.49}
\DefMacro{train-cs-median-new-method-size}{ 122}
\DefMacro{train-cs-mean-old-method-size}{ 200.37}
\DefMacro{train-cs-median-old-method-size}{ 119}
\DefMacro{train-java-mean-add-tokens}{ 19.57}
\DefMacro{train-java-median-add-tokens}{ 4}
\DefMacro{train-java-mean-del-tokens}{ 16.62}
\DefMacro{train-java-median-del-tokens}{ 2}
\DefMacro{train-java-mean-edit-actions}{ 2.71}
\DefMacro{train-java-median-edit-actions}{ 2}
\DefMacro{train-java-mean-edit-lines}{ 5.25}
\DefMacro{train-java-median-edit-lines}{ 2}
\DefMacro{train-java-mean-edit-tokens}{ 36.19}
\DefMacro{train-java-median-edit-tokens}{ 10}
\DefMacro{train-java-mean-new-method-size}{ 195.99}
\DefMacro{train-java-median-new-method-size}{ 117}
\DefMacro{train-java-mean-old-method-size}{ 193.05}
\DefMacro{train-java-median-old-method-size}{ 115}
\DefMacro{valid-cs-mean-add-tokens}{ 17.69}
\DefMacro{valid-cs-median-add-tokens}{ 6}
\DefMacro{valid-cs-mean-del-tokens}{ 17.92}
\DefMacro{valid-cs-median-del-tokens}{ 2}
\DefMacro{valid-cs-mean-edit-actions}{ 2.75}
\DefMacro{valid-cs-median-edit-actions}{ 2}
\DefMacro{valid-cs-mean-edit-lines}{ 6.89}
\DefMacro{valid-cs-median-edit-lines}{ 4}
\DefMacro{valid-cs-mean-edit-tokens}{ 35.61}
\DefMacro{valid-cs-median-edit-tokens}{ 13}
\DefMacro{valid-cs-mean-new-method-size}{ 199.47}
\DefMacro{valid-cs-median-new-method-size}{ 142}
\DefMacro{valid-cs-mean-old-method-size}{ 199.71}
\DefMacro{valid-cs-median-old-method-size}{ 141}
\DefMacro{valid-java-mean-add-tokens}{ 16.64}
\DefMacro{valid-java-median-add-tokens}{ 5}
\DefMacro{valid-java-mean-del-tokens}{ 17.16}
\DefMacro{valid-java-median-del-tokens}{ 2}
\DefMacro{valid-java-mean-edit-actions}{ 2.68}
\DefMacro{valid-java-median-edit-actions}{ 2}
\DefMacro{valid-java-mean-edit-lines}{ 5.50}
\DefMacro{valid-java-median-edit-lines}{ 3}
\DefMacro{valid-java-mean-edit-tokens}{ 33.80}
\DefMacro{valid-java-median-edit-tokens}{ 12}
\DefMacro{valid-java-mean-new-method-size}{ 192.36}
\DefMacro{valid-java-median-new-method-size}{ 134}
\DefMacro{valid-java-mean-old-method-size}{ 192.88}
\DefMacro{valid-java-median-old-method-size}{ 135}

\DefMacro{res-java2cs-cp-copy-Acc-top-1}{0.00}
\DefMacro{res-java2cs-cp-copy-BLEU}{79.96}
\DefMacro{res-java2cs-cp-copy-COPY-PCT}{100.00}
\DefMacro{res-java2cs-cp-copy-CodeBLEU}{89.08}
\DefMacro{res-java2cs-cp-copy-GLEU}{71.89}
\DefMacro{res-java2cs-cp-copy-SARI}{30.53}
\DefMacro{res-java2cs-cp-copy-dataflow_match_score}{81.80}
\DefMacro{res-java2cs-cp-copy-ngram_match_score}{91.39}
\DefMacro{res-java2cs-cp-copy-syntax_match_score}{91.74}
\DefMacro{res-java2cs-cp-copy-weighted_ngram_match_score}{91.39}
\DefMacro{res-java2cs-cp-CopyEdits-Acc-top-1}{14.19}
\DefMacro{res-java2cs-cp-CopyEdits-BLEU}{87.95}
\DefMacro{res-java2cs-cp-CopyEdits-COPY-PCT}{3.22}
\DefMacro{res-java2cs-cp-CopyEdits-CodeBLEU}{89.73}
\DefMacro{res-java2cs-cp-CopyEdits-GLEU}{85.55}
\DefMacro{res-java2cs-cp-CopyEdits-SARI}{67.58}
\DefMacro{res-java2cs-cp-CopyEdits-dataflow_match_score}{80.39}
\DefMacro{res-java2cs-cp-CopyEdits-ngram_match_score}{92.07}
\DefMacro{res-java2cs-cp-CopyEdits-syntax_match_score}{92.06}
\DefMacro{res-java2cs-cp-CopyEdits-weighted_ngram_match_score}{94.41}
\DefMacro{res-java2cs-cp-translation-CodeT5-Acc-top-1}{10.64}
\DefMacro{res-java2cs-cp-translation-CodeT5-BLEU}{77.34}
\DefMacro{res-java2cs-cp-translation-CodeT5-COPY-PCT}{0.00}
\DefMacro{res-java2cs-cp-translation-CodeT5-CodeBLEU}{64.35}
\DefMacro{res-java2cs-cp-translation-CodeT5-GLEU}{74.16}
\DefMacro{res-java2cs-cp-translation-CodeT5-SARI}{71.73}
\DefMacro{res-java2cs-cp-translation-CodeT5-dataflow_match_score}{53.53}
\DefMacro{res-java2cs-cp-translation-CodeT5-ngram_match_score}{63.11}
\DefMacro{res-java2cs-cp-translation-CodeT5-syntax_match_score}{75.28}
\DefMacro{res-java2cs-cp-translation-CodeT5-weighted_ngram_match_score}{65.50}
\DefMacro{res-java2cs-cp-CoditT5-Acc-top-1}{34.59}
\DefMacro{res-java2cs-cp-CoditT5-BLEU}{81.59}
\DefMacro{res-java2cs-cp-CoditT5-COPY-PCT}{0.00}
\DefMacro{res-java2cs-cp-CoditT5-CodeBLEU}{65.17}
\DefMacro{res-java2cs-cp-CoditT5-GLEU}{80.91}
\DefMacro{res-java2cs-cp-CoditT5-SARI}{83.29}
\DefMacro{res-java2cs-cp-CoditT5-dataflow_match_score}{55.65}
\DefMacro{res-java2cs-cp-CoditT5-ngram_match_score}{59.79}
\DefMacro{res-java2cs-cp-CoditT5-syntax_match_score}{80.23}
\DefMacro{res-java2cs-cp-CoditT5-weighted_ngram_match_score}{64.99}
\DefMacro{res-java2cs-cp-update-CodeT5-Acc-top-1}{29.38}
\DefMacro{res-java2cs-cp-update-CodeT5-BLEU}{80.56}
\DefMacro{res-java2cs-cp-update-CodeT5-COPY-PCT}{3.33}
\DefMacro{res-java2cs-cp-update-CodeT5-CodeBLEU}{66.15}
\DefMacro{res-java2cs-cp-update-CodeT5-GLEU}{79.65}
\DefMacro{res-java2cs-cp-update-CodeT5-SARI}{64.70}
\DefMacro{res-java2cs-cp-update-CodeT5-dataflow_match_score}{56.46}
\DefMacro{res-java2cs-cp-update-CodeT5-ngram_match_score}{62.20}
\DefMacro{res-java2cs-cp-update-CodeT5-syntax_match_score}{80.26}
\DefMacro{res-java2cs-cp-update-CodeT5-weighted_ngram_match_score}{65.67}
\DefMacro{res-java2cs-cp-chatgpt3.5-zero-shot-Acc-top-1}{39.58}
\DefMacro{res-java2cs-cp-chatgpt3.5-zero-shot-BLEU}{86.97}
\DefMacro{res-java2cs-cp-chatgpt3.5-zero-shot-COPY-PCT}{2.77}
\DefMacro{res-java2cs-cp-chatgpt3.5-zero-shot-CodeBLEU}{74.90}
\DefMacro{res-java2cs-cp-chatgpt3.5-zero-shot-GLEU}{86.14}
\DefMacro{res-java2cs-cp-chatgpt3.5-zero-shot-SARI}{70.15}
\DefMacro{res-java2cs-cp-chatgpt3.5-zero-shot-dataflow_match_score}{65.44}
\DefMacro{res-java2cs-cp-chatgpt3.5-zero-shot-ngram_match_score}{73.28}
\DefMacro{res-java2cs-cp-chatgpt3.5-zero-shot-syntax_match_score}{85.65}
\DefMacro{res-java2cs-cp-chatgpt3.5-zero-shot-weighted_ngram_match_score}{75.25}
\DefMacro{res-java2cs-cp-metaEdits-Acc-top-1}{38.36}
\DefMacro{res-java2cs-cp-metaEdits-BLEU}{90.79}
\DefMacro{res-java2cs-cp-metaEdits-COPY-PCT}{1.00}
\DefMacro{res-java2cs-cp-metaEdits-CodeBLEU}{91.60}
\DefMacro{res-java2cs-cp-metaEdits-GLEU}{88.94}
\DefMacro{res-java2cs-cp-metaEdits-SARI}{73.45}
\DefMacro{res-java2cs-cp-metaEdits-dataflow_match_score}{84.72}
\DefMacro{res-java2cs-cp-metaEdits-ngram_match_score}{93.64}
\DefMacro{res-java2cs-cp-metaEdits-syntax_match_score}{92.91}
\DefMacro{res-java2cs-cp-metaEdits-weighted_ngram_match_score}{95.13}
\DefMacro{res-java2cs-cp-edit-translation-Acc-top-1}{41.91}
\DefMacro{res-java2cs-cp-edit-translation-BLEU}{90.86}
\DefMacro{res-java2cs-cp-edit-translation-COPY-PCT}{0.44}
\DefMacro{res-java2cs-cp-edit-translation-CodeBLEU}{91.35}
\DefMacro{res-java2cs-cp-edit-translation-GLEU}{88.94}
\DefMacro{res-java2cs-cp-edit-translation-SARI}{74.59}
\DefMacro{res-java2cs-cp-edit-translation-dataflow_match_score}{86.93}
\DefMacro{res-java2cs-cp-edit-translation-ngram_match_score}{88.08}
\DefMacro{res-java2cs-cp-edit-translation-syntax_match_score}{94.05}
\DefMacro{res-java2cs-cp-edit-translation-weighted_ngram_match_score}{96.34}
\DefMacro{res-java2cs-cp-hybrid-Acc-top-1}{43.35}
\DefMacro{res-java2cs-cp-hybrid-BLEU}{92.51}
\DefMacro{res-java2cs-cp-hybrid-COPY-PCT}{0.00}
\DefMacro{res-java2cs-cp-hybrid-CodeBLEU}{91.70}
\DefMacro{res-java2cs-cp-hybrid-GLEU}{91.18}
\DefMacro{res-java2cs-cp-hybrid-SARI}{89.13}
\DefMacro{res-java2cs-cp-hybrid-dataflow_match_score}{87.07}
\DefMacro{res-java2cs-cp-hybrid-ngram_match_score}{88.87}
\DefMacro{res-java2cs-cp-hybrid-syntax_match_score}{94.46}
\DefMacro{res-java2cs-cp-hybrid-weighted_ngram_match_score}{96.41}

\DefMacro{res-cs2java-cp-copy-Acc-top-1}{0.00}
\DefMacro{res-cs2java-cp-copy-BLEU}{80.02}
\DefMacro{res-cs2java-cp-copy-COPY-PCT}{100.00}
\DefMacro{res-cs2java-cp-copy-CodeBLEU}{88.60}
\DefMacro{res-cs2java-cp-copy-GLEU}{71.94}
\DefMacro{res-cs2java-cp-copy-SARI}{30.51}
\DefMacro{res-cs2java-cp-copy-dataflow_match_score}{79.89}
\DefMacro{res-cs2java-cp-copy-ngram_match_score}{91.48}
\DefMacro{res-cs2java-cp-copy-syntax_match_score}{91.55}
\DefMacro{res-cs2java-cp-copy-weighted_ngram_match_score}{91.47}
\DefMacro{res-cs2java-cp-CopyEdits-Acc-top-1}{13.86}
\DefMacro{res-cs2java-cp-CopyEdits-BLEU}{86.99}
\DefMacro{res-cs2java-cp-CopyEdits-COPY-PCT}{2.77}
\DefMacro{res-cs2java-cp-CopyEdits-CodeBLEU}{88.70}
\DefMacro{res-cs2java-cp-CopyEdits-GLEU}{84.14}
\DefMacro{res-cs2java-cp-CopyEdits-SARI}{66.50}
\DefMacro{res-cs2java-cp-CopyEdits-dataflow_match_score}{78.28}
\DefMacro{res-cs2java-cp-CopyEdits-ngram_match_score}{90.81}
\DefMacro{res-cs2java-cp-CopyEdits-syntax_match_score}{91.72}
\DefMacro{res-cs2java-cp-CopyEdits-weighted_ngram_match_score}{94.01}
\DefMacro{res-cs2java-cp-translation-CodeT5-Acc-top-1}{6.21}
\DefMacro{res-cs2java-cp-translation-CodeT5-BLEU}{76.84}
\DefMacro{res-cs2java-cp-translation-CodeT5-COPY-PCT}{0.00}
\DefMacro{res-cs2java-cp-translation-CodeT5-CodeBLEU}{62.27}
\DefMacro{res-cs2java-cp-translation-CodeT5-GLEU}{69.81}
\DefMacro{res-cs2java-cp-translation-CodeT5-SARI}{67.37}
\DefMacro{res-cs2java-cp-translation-CodeT5-dataflow_match_score}{46.91}
\DefMacro{res-cs2java-cp-translation-CodeT5-ngram_match_score}{62.76}
\DefMacro{res-cs2java-cp-translation-CodeT5-syntax_match_score}{74.39}
\DefMacro{res-cs2java-cp-translation-CodeT5-weighted_ngram_match_score}{65.01}
\DefMacro{res-cs2java-cp-CoditT5-Acc-top-1}{35.81}
\DefMacro{res-cs2java-cp-CoditT5-BLEU}{82.89}
\DefMacro{res-cs2java-cp-CoditT5-COPY-PCT}{0.00}
\DefMacro{res-cs2java-cp-CoditT5-CodeBLEU}{65.20}
\DefMacro{res-cs2java-cp-CoditT5-GLEU}{81.67}
\DefMacro{res-cs2java-cp-CoditT5-SARI}{83.27}
\DefMacro{res-cs2java-cp-CoditT5-dataflow_match_score}{53.98}
\DefMacro{res-cs2java-cp-CoditT5-ngram_match_score}{61.22}
\DefMacro{res-cs2java-cp-CoditT5-syntax_match_score}{79.70}
\DefMacro{res-cs2java-cp-CoditT5-weighted_ngram_match_score}{65.90}
\DefMacro{res-cs2java-cp-update-CodeT5-Acc-top-1}{31.60}
\DefMacro{res-cs2java-cp-update-CodeT5-BLEU}{81.98}
\DefMacro{res-cs2java-cp-update-CodeT5-COPY-PCT}{3.66}
\DefMacro{res-cs2java-cp-update-CodeT5-CodeBLEU}{65.82}
\DefMacro{res-cs2java-cp-update-CodeT5-GLEU}{81.07}
\DefMacro{res-cs2java-cp-update-CodeT5-SARI}{65.49}
\DefMacro{res-cs2java-cp-update-CodeT5-dataflow_match_score}{54.26}
\DefMacro{res-cs2java-cp-update-CodeT5-ngram_match_score}{62.84}
\DefMacro{res-cs2java-cp-update-CodeT5-syntax_match_score}{80.24}
\DefMacro{res-cs2java-cp-update-CodeT5-weighted_ngram_match_score}{65.93}
\DefMacro{res-cs2java-cp-chatgpt3.5-zero-shot-Acc-top-1}{39.80}
\DefMacro{res-cs2java-cp-chatgpt3.5-zero-shot-BLEU}{89.35}
\DefMacro{res-cs2java-cp-chatgpt3.5-zero-shot-COPY-PCT}{3.10}
\DefMacro{res-cs2java-cp-chatgpt3.5-zero-shot-CodeBLEU}{76.69}
\DefMacro{res-cs2java-cp-chatgpt3.5-zero-shot-GLEU}{88.62}
\DefMacro{res-cs2java-cp-chatgpt3.5-zero-shot-SARI}{72.36}
\DefMacro{res-cs2java-cp-chatgpt3.5-zero-shot-dataflow_match_score}{65.41}
\DefMacro{res-cs2java-cp-chatgpt3.5-zero-shot-ngram_match_score}{76.65}
\DefMacro{res-cs2java-cp-chatgpt3.5-zero-shot-syntax_match_score}{86.72}
\DefMacro{res-cs2java-cp-chatgpt3.5-zero-shot-weighted_ngram_match_score}{77.99}
\DefMacro{res-cs2java-cp-metaEdits-Acc-top-1}{41.91}
\DefMacro{res-cs2java-cp-metaEdits-BLEU}{91.54}
\DefMacro{res-cs2java-cp-metaEdits-COPY-PCT}{0.89}
\DefMacro{res-cs2java-cp-metaEdits-CodeBLEU}{91.21}
\DefMacro{res-cs2java-cp-metaEdits-GLEU}{89.36}
\DefMacro{res-cs2java-cp-metaEdits-SARI}{73.46}
\DefMacro{res-cs2java-cp-metaEdits-dataflow_match_score}{82.72}
\DefMacro{res-cs2java-cp-metaEdits-ngram_match_score}{93.61}
\DefMacro{res-cs2java-cp-metaEdits-syntax_match_score}{92.94}
\DefMacro{res-cs2java-cp-metaEdits-weighted_ngram_match_score}{95.57}
\DefMacro{res-cs2java-cp-edit-translation-Acc-top-1}{40.35}
\DefMacro{res-cs2java-cp-edit-translation-BLEU}{91.63}
\DefMacro{res-cs2java-cp-edit-translation-COPY-PCT}{0.55}
\DefMacro{res-cs2java-cp-edit-translation-CodeBLEU}{90.99}
\DefMacro{res-cs2java-cp-edit-translation-GLEU}{89.58}
\DefMacro{res-cs2java-cp-edit-translation-SARI}{74.15}
\DefMacro{res-cs2java-cp-edit-translation-dataflow_match_score}{83.73}
\DefMacro{res-cs2java-cp-edit-translation-ngram_match_score}{90.19}
\DefMacro{res-cs2java-cp-edit-translation-syntax_match_score}{93.90}
\DefMacro{res-cs2java-cp-edit-translation-weighted_ngram_match_score}{96.14}
\DefMacro{res-cs2java-cp-hybrid-Acc-top-1}{46.34}
\DefMacro{res-cs2java-cp-hybrid-BLEU}{93.17}
\DefMacro{res-cs2java-cp-hybrid-COPY-PCT}{0.00}
\DefMacro{res-cs2java-cp-hybrid-CodeBLEU}{91.32}
\DefMacro{res-cs2java-cp-hybrid-GLEU}{91.24}
\DefMacro{res-cs2java-cp-hybrid-SARI}{89.62}
\DefMacro{res-cs2java-cp-hybrid-dataflow_match_score}{84.16}
\DefMacro{res-cs2java-cp-hybrid-ngram_match_score}{90.58}
\DefMacro{res-cs2java-cp-hybrid-syntax_match_score}{94.30}
\DefMacro{res-cs2java-cp-hybrid-weighted_ngram_match_score}{96.26}

\DefMacro{FCap-model-figure}{Workflow of \DeltaTool for \deltatask.
\DeltaTool leverages the context of code change histories of multiple
programming languages from three sources: code changes on the source
programming language (\UseMacro{java_edits}), the old version of code
in the target programming language (\UseMacro{cs_old}), and the new
version of code in the source programming language
(\UseMacro{java_new}). \DeltaTool has two variants that both generate
the code changes in the target programming language
(\UseMacro{cs_edits}) but in different formats: \edittranslation
directly generates the code changes; \metaEdits generates the meta
edit plan which edits \UseMacro{java_edits} to \UseMacro{cs_edits},
followed by the code changes. Finally, we apply the code changes
(\UseMacro{cs_edits}) on the old version of code (\UseMacro{cs_old})
to obtain the new version of code (\UseMacro{cs_new}) in the target
programming language.  \label{fig:model-architecture}}
\DefMacro{FCap-xMatch-tokens-number}{Average percentage of model's predictions that exactly match the ground truth on examples that have different number of subtokens. The bands represent the 95\% confidence interval.}
\DefMacro{FCap-ref-tokens-number-dist}{Distribution of number of sub tokens in models' target outputs.}

\DefMacro{intro-example-java-method}{\texttt{doc\-With\-Invalid\-Mapp\-ing02}}
\DefMacro{intro-example-cs-method}{\texttt{Doc\-With\-Invalid\-Mapp\-ing02}}
\DefMacro{intro-example-old-edits}{\texttt{<ReplaceOld> PdfException <ReplaceNew> LayoutExceptionMessageConstant <ReplaceEnd>}}
\DefMacro{intro-example-java-edits}{\texttt{<ReplaceOldKeepBefore> format(PdfException}\\\texttt{<ReplaceNewKeepBefore> format(}\\\texttt{LayoutExceptionMessageConstant <ReplaceEnd>}}
\DefMacro{intro-example-cs-edits}{\texttt{<ReplaceOldKeepBefore> Format(PdfException}\\\texttt{<ReplaceNewKeepBefore> Format(}\\\texttt{LayoutExceptionMessageConstant <ReplaceEnd>}}
\DefMacro{intro-example-meta-edits}{\texttt{<ReplaceOld> format <ReplaceNew> Format <ReplaceEnd>}}
\DefMacro{intro-example-java-concise-edits}{\texttt{<ReplaceOld> PdfException <Replace\-New> LayoutExceptionMessageConstant <ReplaceEnd>}}

\DefMacro{git-add-color}{green!10}
\DefMacro{git-del-color}{red!10}

\title{\Title}

\author{Jiyang Zhang}
\affiliation{
  \institution{UT Austin}
  \country{USA}
}
\email{jiyang.zhang@utexas.edu}

\author{Pengyu Nie}
\affiliation{
  \institution{UT Austin}
  \country{USA}
}
\email{pynie@utexas.edu}

\author{Junyi Jessy Li}
\affiliation{
  \institution{UT Austin}
  \country{USA}
}
\email{jessy@austin.utexas.edu}

\author{Milos Gligoric}
\affiliation{
  \institution{UT Austin}
  \country{USA}
}
\email{gligoric@utexas.edu}

\begin{document}

\begin{abstract}
Many software projects implement \APIs and algorithms in multiple
programming languages.  Maintaining such projects is tiresome, as
developers have to ensure that any change (e.g., a bug fix or a new
feature) is being propagated, timely and without errors, to
implementations in other programming languages.
In the world of ever-changing software, using rule-based translation
tools (i.e., transpilers) or machine learning models for translating
code from one language to another provides limited value.  Translating
each time the entire codebase from one language to another is not the
way developers work.
In this paper, we target a novel task: translating code changes from
one programming language to another using large language models
(\LLMs).
We design and implement the first \LLM, dubbed \DeltaTool, to
tackle this task.  \DeltaTool explicitly models code changes as edit
sequences and learns to correlate changes across programming languages.
To evaluate \DeltaTool, we collect a corpus of \Datasize{} aligned
code changes from \ProjectsNum{} pairs of open-source software
projects implementing similar functionalities in two programming
languages (Java and \Cs).  Results show that \DeltaTool
outperforms the state-of-the-art approaches by a large margin on all
commonly used automatic metrics.
Our work also reveals that \DeltaTool is complementary to the
existing generation-based models, and their combination
ensures even greater performance.
\end{abstract}

\vspace{-10pt}


\maketitle

\section{Introduction}

To ensure flexibility and a wide adoption of their software, companies
provide application programming interfaces (\APIs) for their services
in several programming languages.  Services, such as Google
Cloud~\cite{GoogleCloud} and MongoDB~\cite{MongoDB}, offer APIs
written in most popular programming languages, including C++, C\#,
Java, and Python.  Furthermore, popular software packages, like
Antlr~\cite{antlr} and Lucene~\cite{Lucene}, have options to target
different programming languages for the purpose of being used across
various platforms easily.

Maintaining software that offers the same functionality in multiple
programming languages is challenging.  Any code change, due to a
feature request or a bug fix, has to be propagated timely to all
programming languages.  At present, developers have to manually
\emph{co-evolve} code.  This requires developers to manually find the
correspondence between code snippets and apply necessary \emph{edits}.
\begin{figure}[t]
  \centering
  \vspace{5pt}
  \newsavebox\boxJavaExample
\begin{lrbox}{\boxJavaExample}
  \begin{lstlisting}[language=java-pretty, numbers=left]
      @Test
      public void docWithInvalidMapping02() throws IOException {
        ...
        customRolePara.getAccessibilityProperties().setRole(HtmlRoles.p);
        Exception e = Assert.assertThrows(PdfException.class,()->document.add(customRolePara));
       - Assert.assertEquals(MessageFormat.format((-w<)PdfException(>w-).ROLE_IS_NOT_MAPPED_TO_ANY_STANDARD_ROLE, "p"), e.getMessage());
       + Assert.assertEquals(MessageFormat.format((+w<)LayoutExceptionMessageConstant(>w+).ROLE_IS_NOT_MAPPED_TO_ANY_STANDARD_ROLE, "p"), e.getMessage());
      }
  \end{lstlisting}

\end{lrbox}

\newsavebox\boxGenerationExample
\begin{lrbox}{\boxGenerationExample}
  \begin{lstlisting}[language=java-pretty, numbers=left]
      [NUnit.Framework.Test] 
      public virtual void DocWithInvalidMapping02() {
        ...
        - customRolePara.GetAccessibilityProperties().SetRole((-w<)LayoutTaggingPdf2Test(>w-).HtmlRoles.p);
        + customRolePara.GetAccessibilityProperties().SetRole(HtmlRoles.p);
        - Exception e = NUnit.Framework.Assert.(-w<)Catch(>w-)(typeof(PdfException),()=>document.Add(customRolePara));
        + Exception e = NUnit.Framework.Assert.(+w<)IsThrows(>w+)(PdfException.class,()=>document.Add(customRolePara));
        - NUnit.Framework.Assert.AreEqual(String.Format((-w<)PdfException(>w-).ROLE_IS_NOT_MAPPED_TO_ANY_STANDARD_ROLE, "p"), e.Message);
        + NUnit.Framework.Assert.AreEqual(String.Format((+w<)LayoutExceptionMessageConstant(>w+).ROLE_IS_NOT_MAPPED_TO_ANY_STANDARD_ROLE, "p"), e.Message);
      }
  \end{lstlisting}
\end{lrbox}

\newsavebox\boxCsOurExample
\begin{lrbox}{\boxCsOurExample}
  \begin{lstlisting}[language=java-pretty, numbers=left]
      [NUnit.Framework.Test] 
      public virtual void DocWithInvalidMapping02() {
        ...
        customRolePara.GetAccessibilityProperties().SetRole(
          LayoutTaggingPdf2Test.HtmlRoles.p);
        Exception e = NUnit.Framework.Assert.Catch(typeof(PdfException),()=>document.Add(customRolePara));
        - NUnit.Framework.Assert.AreEqual(String.Format((-w<)PdfException(>w-).ROLE_IS_NOT_MAPPED_TO_ANY_STANDARD_ROLE, "p"), e.Message);
        + NUnit.Framework.Assert.AreEqual(String.Format((+w<)LayoutExceptionMessageConstant(>w+).ROLE_IS_NOT_MAPPED_TO_ANY_STANDARD_ROLE, "p"), e.Message);
      }
  \end{lstlisting}
\end{lrbox}


\newsavebox\boxJavaAdd
\begin{lrbox}{\boxJavaAdd}
  \begin{lstlisting}[language=java-pretty]
    
  \end{lstlisting}
\end{lrbox}

\begin{adjustbox}{margin=-0.2cm 0cm 0cm 0cm}
\begin{tikzpicture}[
    line width=0.4pt,
    node distance=0ex and 0em,
    every node/.style={scale=1},
    gridBox/.style={rectangle, opacity=0, draw=red},
    box/.style={rectangle, draw=black, inner sep=2pt, font=\small},
    rounded box/.style={rectangle, rounded corners, draw=black, inner sep=2pt, font=\small},
    anno/.style={font=\footnotesize},
]

    \DefMacro{wCodeBox}{26.1em}
    \DefMacro{hJavaCodeBox}{25ex}
    \DefMacro{hGenerationCodeBox}{35ex}
    \DefMacro{hOurCodeBox}{29ex}

    \DefMacro{diff-box-w}{25.9em}
    
    \node (box-java) at (0,0) [box, minimum width=\UseMacro{wCodeBox}, minimum height=\UseMacro{hJavaCodeBox}] {};
    \node (java-del) [below right = 13.5ex and .1em of box-java.north west] [box, minimum width=\UseMacro{diff-box-w}, minimum height=4ex, draw=none, fill=\UseMacro{git-del-color}, ] {};
    \node (java-add) [below right = 17.5ex and .1em of box-java.north west] [box, minimum width=\UseMacro{diff-box-w}, minimum height=4ex, draw=none, fill=green!20, ] {};
    \node (b-java) [below right = 0 and -1em of box-java.north west] [] {\usebox\boxJavaExample};
    \node (b-java-text) [above left = .5ex and .1em of box-java.south east] [box, scale=0.8] {\bf Java Change Made by Developers\phantom{p\hskip -.6em}};

    \node (box-generation) [below = 0 of box-java.south] [box, minimum width=\UseMacro{wCodeBox}, minimum height=\UseMacro{hGenerationCodeBox}] {};
    \node (gen-del-0) [below right = 7.5ex and .1em of box-generation.north west] [box, minimum width=\UseMacro{diff-box-w}, minimum height=4ex, draw=none, fill=\UseMacro{git-del-color}, ] {};
    \node (gen-add-0) [below right = 11.5ex and .1em of box-generation.north west] [box, minimum width=\UseMacro{diff-box-w}, minimum height=2ex, draw=none, fill=green!20, ] {};
    \node (gen-del-1) [below right = 13.75ex and .1em of box-generation.north west] [box, minimum width=\UseMacro{diff-box-w}, minimum height=4ex, draw=none, fill=\UseMacro{git-del-color}, ] {};
    \node (gen-add-1) [below right = 17.75ex and .1em of box-generation.north west] [box, minimum width=\UseMacro{diff-box-w}, minimum height=4ex, draw=none, fill=green!20, ] {};
    \node (gen-del-2) [below right = 22ex and .1em of box-generation.north west] [box, minimum width=\UseMacro{diff-box-w}, minimum height=4ex, draw=none, fill=\UseMacro{git-del-color}, ] {};
    \node (gen-add-2) [below right = 26ex and .1em of box-generation.north west] [box, minimum width=\UseMacro{diff-box-w}, minimum height=6ex, draw=none, fill=green!20, ] {};
    \node (generation) [below right = 0 and -1em of box-generation.north west] [] {\usebox\boxGenerationExample};
    \node (b-generation-text) [above left = .5ex and .1em of box-generation.south east] [box, draw,  scale=0.8] {\bf \Cs Change Predicted by Existing Generation-based Model \ \textcolor{red!80!black}{\mycross}\phantom{p\hskip -.6em}};

    \node (box-our) [below = 0 of box-generation.south] [box, minimum width=\UseMacro{wCodeBox}, minimum height=\UseMacro{hOurCodeBox}] {};
    \node (our-del) [below right = 15.75ex and .1em of box-our.north west] [box, minimum width=\UseMacro{diff-box-w}, minimum height=4ex, draw=none, fill=\UseMacro{git-del-color}, ] {};
    \node (our-add) [below right = 19.75ex and .1em of box-our.north west] [box, minimum width=\UseMacro{diff-box-w}, minimum height=6ex, draw=none, fill=green!20, ] {};
    \node (our) [below right = 0 and -1em of box-our.north west] [] {\usebox\boxCsOurExample};
    \node (b-our-text) [above left = .5ex and .1em of box-our.south east] [box, draw,  scale=0.8] {\bf \Cs Change Made by Developers and Predicted by Our \DeltaTool{} \ \textcolor{green!80!black}{\mycheckmark}\phantom{p\hskip -.6em}};
    
\end{tikzpicture}
\end{adjustbox}
    \vspace{-10pt}
\caption{Example of using LLMs to help developers co-evolve code in
two programming languages.  The top box shows developer-made changes
in a \Java{} project \CodeIn{itext/itext7}, which needs to be
propagated to the corresponding \Cs{} project
\CodeIn{itext/itext7\text{-}dotnet}. The middle box shows the prediction by
an existing generation-based large language model, which incorrectly
changes irrelevant parts of the code. The bottom box shows the correct
prediction by our model, \DeltaTool. \label{fig:intro-example}\vspace{-9pt}}
\vspace{-9pt}
\end{figure}

There has been work that could, in theory, help with translation.
Rule-based migration tools~\cite{java2csharp, c2rust, j2eif} have been
designed to translate between high-level programming languages (e.g.,
\Java{} and \Cs{}).  However, rule-based systems require developers
who have expertise with both programming languages to manually write
rules to specify the translation mappings.  And the rules need to be
updated with the evolution of programming languages themselves; they
quickly become outdated~\cite{BuiETAL18Hierarchical, java2csharp}.
Recent work on automatic code translation~\cite{RoziereETAL21TransCoderST,
  CodeXGLUE, TipirneniETAL22StructCoder, LachauxETAL20TransCoder, ZhuETAL22Multilingual} aim to directly
translate between a source and a target programming language with the
help of \LLMs, which are 
\pretrained on multiple programming languages.  While these techniques
could be used to produce code snippets that look correct, they make
irrelevant changes that deviate substantially from the newly
introduced features in the source programming language, or they fail
to precisely infer the project-specific data types and class names.

Figure~\ref{fig:intro-example} illustrates the limitation of existing
models.  Developers changed \CodeIn{PdfException} to
\CodeIn{LayoutExceptionMessageConstant} in method
\UseMacro{intro-example-java-method} in the \Java project
\CodeIn{itext/}\-\CodeIn{itext7}.
In a later commit in the corresponding \Cs project
\CodeIn{itext/}\-\CodeIn{itext7\text{-}dotnet}, developers revised method
\UseMacro{intro-example-cs-method} with exactly the same edits while
keeping other parts of the method unchanged.
We provide the \Java code change, the prediction of an existing large
language model, \CodeTf~\cite{WangETAL21codet5}, \finetuned for code
translation, and the correct \Cs code change in
Figure~\ref{fig:intro-example}.  The added lines of code are
highlighted in green and the removed ones are highlighted in red.
Although the existing model is able to correctly translate the updated
exception type from \Java to \Cs, it misses the class name for the
field \CodeIn{HtmlRoles} and incorrectly infers the function call
\CodeIn{Assert.Catch} as it does \emph{not} use the prior version of
\Cs code for reference.

To build more robust and accurate techniques that help software
developers co-evolve projects implemented in different languages, we
explicitly model the \emph{changes} that need to be made.
We formulate a novel task:
automatically \emph{updating} code snippets in a target programming
language, based on the \emph{changes} made in the source programming
language.

Most of the existing models implicitly tackle the code evolution tasks
by generating tokens one by one in accordance with the underlying
learned probability instead of focusing on how the code should be
\emph{modified} or retained.  Prior work~\cite{ZhangETAL22CoditT5,
YaoETAL21Learning, ChakrabortyAndRay21Multi, Codit,
TufanoETAL19Learning, DingETAL20Patching, PanthaplackelETAL20Learning}
have shown that standard generation-based models underperform models
that explicitly model the edits on software-editing tasks.

To model code evolution across programming languages, we design an
LLM, dubbed \DeltaTool, which learns to align the edits across
programming languages and explicitly performs edits on the old version
of the code in a target programming language.
Following prior work~\cite{ZhangETAL22CoditT5,
  PanthaplackelETAL20Learning, DingETAL20Patching, stahlberg2020seq2edits}, we
enable the model to reason about necessary 
edits and learn to apply them by directly generating an edit sequence.

For training and evaluation, we collect the first dataset with aligned
\Java and \Cs code changes on the methods with similar functionality
and implementations.  Specifically, we extract \Datasize{} pairs of
code changes from \ProjectsNum{} open-source \Java{} projects and the
corresponding \Cs{} projects on GitHub by mining the commit histories.
This is the first dataset containing parallel code changes of two
programming languages.
We conduct the evaluation in two directions, updating \Cs method based
on the \Java changes (source language is \Java and target language is
\Cs) and updating \Java method based on the \Cs changes (source
language is \Cs and target language is \Java).

Our results show that \DeltaTool outperforms all existing models
across all the chosen automatic metrics, including the large \pretrained
generative models \Codex~\cite{Codex} under few-shot setting and \chatgpt~\cite{OpenAIChatGPT} under zero-shot setting.  \DeltaTool achieves
\BestCodeBLEU{} (out of 100) CodeBLEU score on the task of updating \Cs
code based on \Java changes, which is more than
\CodeBLEUImproved{} higher than the large \pretrained generation-based
model \finetuned on this task.

Further, we find that \DeltaTool and generation-based models are
complementary to each other as \DeltaTool is better at updating longer
code snippets while generation model is better at handling the shorter
ones.  Thus, we combine the two models by choosing either model's
prediction based on the size of the input code.
Our results show that the combination can further improve our
\DeltaTool model's exact-match accuracy by 6\%.

\vspace{3pt}
\noindent
The main contributions of this paper include:

\begin{itemize}[topsep=3pt,itemsep=1ex,partopsep=0ex,parsep=0ex,leftmargin=*]
\item \textbf{Task}. We formulate a novel task of automatically
updating code written in one programming language based on the changes
in the corresponding code in another programming language.
\item \textbf{Model}. We design and implement \DeltaTool, the first
\LLM for this task which learns to align the edits across programming
languages and explicitly performs edits on the old version of the code
in  target programming language.
\item \textbf{Dataset}. We create the first dataset with aligned code
  changes for two programming languages (Java and C\#) from
  \ProjectsNum{} open-source project pairs.
\item \textbf{Results}. We show that \DeltaTool significantly
outperforms the existing LLMs \finetuned for code translation on exact-match accuracy by 
\BestImproved{}.
We also show
that \DeltaTool is complementary to generation-based \LLMs and the
combination can further improve \DeltaTool's exact-match accuracy by
6\%.
\end{itemize}

\noindent
\DeltaTool and our corpus are publicly available on GitHub:\\
  \url{https://github.com/EngineeringSoftware/codeditor}.

\vspace{-10pt}

\section{Task}

At a high level, we work on a system that is triggered when a software
developer, who maintains projects written in multiple programming
languages, makes changes to one method in one of the languages, i.e.,
the ``source'' language.
The system would automatically suggest updates to the methods with
identical functionality in other language(s), i.e., the ``target''
language(s).
To scope our work in this paper, we focus on \Java as the source
language, and \Cs as the target language.
We leave evaluation that targets other programming languages as future
work.

In Figure~\ref{fig:intro-example}, consider a method
\UseMacro{java_old} (\UseMacro{intro-example-java-method}) written in
the source language \UseMacro{source} and a method \UseMacro{cs_old}
(\UseMacro{intro-example-cs-method}) written in the target language
\UseMacro{target} with identical functionality (hence similar
implementation).  Given the updated method \UseMacro{java_new} in
\UseMacro{source}, we define the task to generate the new method
\UseMacro{cs_new} in \UseMacro{target} leveraging context provided by
the code changes \UseMacro{java_edits}, such that its functionality is
consistent with \UseMacro{java_new}.  Namely, we model the conditional
probability distribution
\begin{center}
$P(\UseMacro{cs_new} | \UseMacro{cs_old},\: \UseMacro{java_new}, \UseMacro{java_edits})$
\end{center}
and generate \UseMacro{cs_new} by sampling
from the distribution.

\begin{figure*}[t]
  \centering
  \definecolor{babyblueeyes}{rgb}{0.63, 0.79, 0.95}
\definecolor{bubbles}{rgb}{0.91, 1.0, 1.0}
\definecolor{lavenderblue}{rgb}{0.8, 0.8, 1.0}
\definecolor{gray(x11gray)}{rgb}{0.75, 0.75, 0.75}
\definecolor{cobalt}{rgb}{0.0, 0.28, 0.67}
\definecolor{navyblue}{rgb}{0.0, 0.0, 0.5}
\definecolor{orange(colorwheel)}{rgb}{1.0, 0.5, 0.0}
\definecolor{lightgray}{rgb}{0.83, 0.83, 0.83}

\begin{tikzpicture}[
    node distance = 5cm,
    every node/.style={scale=1},
    roundnode/.style={circle, draw, thin, minimum size=1em},
]

\tikzset{XOR/.style={draw,circle,append after command={
        [shorten >=\pgflinewidth, shorten <=\pgflinewidth,]
        (\tikzlastnode.north) edge (\tikzlastnode.south)
        (\tikzlastnode.east) edge (\tikzlastnode.west)
        }
    }
}

\tikzset{
    pre/.style={=stealth',semithick},
    post/.style={->,shorten >=1pt,>=stealth',semithick}
}

\node [block, minimum width=3.3cm, minimum height=0.8cm, text width=3.3cm, text height=0.2cm, thin, draw, fill=lavenderblue!50, font=\ttfamily, align=left] at (0,0) (javaEdits) {\scriptsize \texttt{<ReplaceOldKeepBefore> format(PdfException <ReplaceNewKeepBefore> format( LayoutExceptionMessageConstant <ReplaceEnd>}};
\node [text_box, left = -2ex of javaEdits, text width=4em] (input1) {\footnotesize \UseMacro{java_edits}};

\node [block, minimum width=3.3cm, minimum height=0.8cm, text width=3.3cm, text height=0.1cm,thin, draw, fill=bubbles, font=\ttfamily, align=left] at (0,-1.7) (csharpOld) {\scriptsize \texttt{...\\String.Format(PdfException...}};
\node [text_box, left = -1.5ex of csharpOld, text width=4em] (input2) {\footnotesize \UseMacro{cs_old}};

\node [block, minimum width=3.3cm, minimum height=0.8cm, text width=3.3cm,text height=0.2cm,thin, draw, fill=babyblueeyes, font=\ttfamily, align=left] at (0,-3) (javaNew) {\scriptsize \texttt {...\\MessageFormat.format(\\LayoutExceptionMessageConstant..}};
\node [text_box, left = -1.5ex of javaNew, text width=4em] (input3) {\footnotesize \UseMacro{java_new}};

\node (concat) [XOR, scale=1.2] at (2.5,-1.7) {};

\node [block, minimum width=7.5cm, minimum height=3cm, dashed, very thick, draw=navyblue, fill=blue!5!white] at (6.95, 0) (editTranslation) {};
\node [text_box, below right = .5ex and .5em of editTranslation.north west, align=left] {\edittranslation};
\node [block, minimum width=7.5cm, minimum height=3cm, dashed, very thick, draw=orange(colorwheel), fill=yellow!5!white] at (6.95, -3.2) (metaEdits) {};
\node [text_box, above right = .5ex and .5em of metaEdits.south west, align=left] {\metaEdits};

\node [block, draw, thin, fill=lightgray,minimum width=3.3cm, text width=3.3cm, text height=0.2cm, font=\ttfamily, align=left] at (6.75,-0.2) (csharpEdits1) {\scriptsize \texttt{<ReplaceOldKeepBefore> Format(PdfException <ReplaceNewKeepBefore> Format( LayoutExceptionMessageConstant <ReplaceEnd>}};
\node [text_box, above = 0.5ex of csharpEdits1, text width=4em] (l-csharpEdits1) {\footnotesize \UseMacro{cs_edits}};

\node [block, draw, thin, fill=lightgray,minimum width=2.2cm, text width=2.2cm, text height=0.2cm, font=\ttfamily, align=left] at (5,-3.2) (metaEdits) {\scriptsize \texttt{<ReplaceOld> format <ReplaceNew> Format <ReplaceEnd>}};
\node [text_box, above = 0.5ex of metaEdits, text width=12em] (l-metaEdits) {\footnotesize meta edit sequence};

\node [block, draw, thin, fill=lightgray,minimum width=3.3cm, text width=3.3cm, text height=0.2cm, font=\ttfamily, align=left] at (8.5,-3.2) (csharpEdits2) {\scriptsize \texttt{<ReplaceOldKeepBefore> Format(PdfException <ReplaceNewKeepBefore> Format( LayoutExceptionMessageConstant <ReplaceEnd>}};
\node [text_box, above = 0.5ex of csharpEdits2, text width=4em] (l-csharpEdits2) {\footnotesize \UseMacro{cs_edits}};

\node [block, thin, draw, fill=lightgray, minimum width=3.3cm, text width=3.3cm, text height=0.2cm,font=\ttfamily, align=left] at (13,-1.7) (csharpNew) {\scriptsize \texttt{...\\String.Format(\\LayoutExceptionMessageConstant..}};
\node [text_box, above = 0.5ex of csharpNew, text width=4em] (l-csharpNew) {\footnotesize \UseMacro{cs_new}};

\draw [post, rounded corners=5pt] (javaEdits.east) -| (concat.north);
\path [line, semithick] (csharpOld.east) -- (concat.west);
\draw [post, rounded corners=5pt] (javaNew.east) -| (concat.south);

\node [coordinate, right = 1em of concat.east] (c-split) {};
\draw [post, rounded corners=5pt] (concat.east) -- (c-split) |- (csharpEdits1.west);
\draw [post, rounded corners=5pt] (concat.east) -- (c-split) |- (metaEdits.west);
\draw [post, rounded corners=5pt] (metaEdits.east) -- (csharpEdits2.west);

\node [coordinate, left = 1em of csharpNew.west] (c-merge) {};
\draw [post, rounded corners=5pt] (csharpEdits1.east) -| (c-merge) -- (csharpNew.west);
\draw [post, rounded corners=5pt] (csharpEdits2.east) -| (c-merge) -- (csharpNew.west);


\end{tikzpicture}
  \vspace{-10pt}
  \caption{\UseMacro{FCap-model-figure}}
  \vspace{-5pt}
\end{figure*}

\section{Model}

We present the overview of the proposed \DeltaTool model in
Figure~\ref{fig:model-architecture}.  \DeltaTool is built upon the
encoder-decoder framework which consists of a transformer-based
encoder and a transformer-based decoder~\cite{VaswaniETAL17Attention}.
Many conditional generation tasks, including code summarization and
translation, are being addressed with encoder-decoder
models~\cite{WangETAL21codet5, PLBART, GuoETAL22Unixcoder, CodeT5Plus, Nie23Thesis}.

We initialize \DeltaTool's parameters with the \pretrained language
model CoditT5~\cite{ZhangETAL22CoditT5}. CoditT5 has shown promising
results on various software-related editing tasks \emph{in a single
programming language}, but nonetheless would provide us with a
``warm-start'' that carries the necessary inductive biases towards
modeling edits.  To adapt to the \deltatask task, we then
fine-tune the \DeltaTool model exploring two key components: (i)~the
context fed into the model; (ii)~the output format of the model.

To encourage our \DeltaTool model to leverage the (synchronous) code
change histories of multiple programming languages in
its training data, we provide the model with context from three
sources as shown in Figure~\ref{fig:model-architecture}: (i)~code changes on source programming language
(\UseMacro{java_edits});
(ii)~old version of the code written in target programming language
(\UseMacro{cs_old});
(iii)~new version of the code written in source programming language
(\UseMacro{java_new}).

We explore two formats to represent the generated code changes:
(i)~the code edits in the target programming language (\UseMacro{cs_edits}); (ii)~a meta
edit sequence that translates the code edits from the source programming
language to the target programming language, followed by the code
edits in the target programming language (this is similar to the
output format of CoditT5).
\noindent
In both cases, we then apply the generated code edits in the target
programming language (\UseMacro{cs_edits}) to the old version of the
code (\UseMacro{cs_old}) to obtain the new version of the code
(\UseMacro{cs_new}).

\newcommand{\specialtwocell}[2][c]{%
  \begin{tabular}[#1]{@{}l@{}}#2\end{tabular}}

\newcommand{\specialthreecell}[2][c]{%
  \begin{tabular}[#1]{@{}l@{}l@{}}#2\end{tabular}}

\begin{small}
\begin{table}[t]
\begin{center}
\caption{\UseMacro{TCap-edits-map}}
\vspace{-8pt}
\begin{tabular}{l l l}
    \toprule
    \textbf{Edit}
    & \textbf{\Concise}
    & \textbf{\Unambig}
    \\
    \midrule
    \multirow{2}{*}{Insertion}
    & \multirow{2}{*}{\texttt{<Insert>}}
    &
    \specialtwocell[t]{\texttt{<ReplaceKeepBefore>}\\ \texttt{<ReplaceKeepAfter>}}
    \\
    \midrule
    \multirow{3}{*}{Deletion}
    & \multirow{3}{*}{\texttt{<Delete>}}
    & %
    \specialthreecell[t]{\texttt{<Delete>} \\ \texttt{<ReplaceKeepBefore>}\\ \texttt{<ReplaceKeepAfter>}}
    \\
    \midrule
    \multirow{3}{*}{Replacement}
    & \multirow{3}{*}{\texttt{<Replace>}}
    & \specialthreecell[t]{\texttt{<Replace>} \\ \texttt{<ReplaceKeepBefore>}\\ \texttt{<ReplaceKeepAfter>}}
    \\
    \bottomrule
\end{tabular}
\vspace{\UseMacro{TVspace-edits-map}}
\end{center}
\end{table}
\end{small}

\subsection{Edit Representations}
\label{sec:edits}

\subsubsection{\Concise Edit Sequence}
\label{sec:edits-sequence}
We first represent edits using a sequence of edits identical to that
used in CoditT5~\cite{ZhangETAL22CoditT5}, which we call \concise edit
sequence.  Each edit is represented as:
\MyQuote{
\texttt{<Operation> [token span] <OperationEnd>}
}
Here, \texttt{<Operation>} is either \CodeIn{Insert}, \CodeIn{Delete}
or \CodeIn{Replace}.  Note that the \CodeIn{Replace} is
represented in a slightly different structure since we must specify
both the old contents to be replaced and the new contents to replace
with:
\MyQuote{
\texttt{<ReplaceOld> [old contents] <ReplaceNew>}\\
\texttt{[new contents] <ReplaceEnd>}
}
For example, in Figure~\ref{fig:intro-example}, the code change on the
old \Java method can be represented by
``\UseMacro{intro-example-java-concise-edits}''.

We use \texttt{difflib}~\cite{difflib} to compute the set of minimal edit sequence from the old and new versions of code.

\subsubsection{\Unambig Edit Sequence}
\label{sec: unambiguous-edits}
One drawback of CoditT5~\cite{ZhangETAL22CoditT5}'s representation
specified above is that the \concise edit sequence can be ambiguous
due to the absence of positional information.  For example, the
\Java{} code change in Figure~\ref{fig:intro-example} can be
represented using \CodeIn{Replace} as:
``\UseMacro{intro-example-java-concise-edits}''.  Without further
specification, the edit does not contain any clues regarding which
\CodeIn{PdfException} should be replaced as there are two occurrences
of \CodeIn{PdfException} in the old code sequence.
For similar reasons, \CodeIn{Insert} is always ambiguous
because of not indicating where to add the new contents and
\CodeIn{Delete} is ambiguous in cases where multiple
occurrences of token spans can be removed.

To eliminate the potential ambiguity in the \concise edit sequence, we
design the format of \unambig edit sequence by adjusting the condensed
edit sequence proposed by~\citet{PanthaplackelETAL20Learning}, which
uses anchor tokens to specify the location to perform
edits.

\MyPara{Insertion}
We do not use \CodeIn{Insert} since it will always
introduce ambiguity without location information.  To represent
insertion, we first find unique anchor tokens that are the shortest
span of tokens that is either before or after the edit location and is
unique in the input sequence.  Then we use \CodeIn{ReplaceKeepBefore}
or \CodeIn{ReplaceKeep}\-\CodeIn{After}, which represents replacing the anchor tokens with
the inserted contents and the anchor tokens.
For example, in Figure~\ref{fig:intro-example}, suppose the \Java{}
code change entails adding a blank return statement after the
\CodeIn{assertEquals} statement on line \JavaInsertLineNumber.
The token span ``\CodeIn{getMessage());}'' will serve as the minimal span
of anchor tokens because it is unique among the old \Java{} code
sequence, and it occurs right before the edit to be performed.  We
disambiguate the edit sequence:
\MyQuote{
\texttt{<Insert> return; <InsertEnd>}
}
with the \unambig edit sequence:
\MyQuote{
\texttt{<ReplaceOldKeepBefore> getMessage());}\\
\texttt{<ReplaceNewKeepBefore> getMessage()); return;}\\
\texttt{<ReplaceEnd>}
}
This edit sequence indicates that ``\CodeIn{getMessage());}'' should
be replaced with ``\CodeIn{getMessage()); return;}''.
We introduce \CodeIn{ReplaceKeep}\-\CodeIn{Before} where the
tokens that follows the \texttt{<ReplaceOldKeep\-Before>} should be
removed and the tokens following
\texttt{<ReplaceNew\-Keep\-Before>} should be inserted.  Different from
\CodeIn{Replace}, there is some overlap between the tokens to be
removed and tokens to be inserted.
If anchor tokens do not exist before the edit location, we use
\CodeIn{ReplaceKeepAfter} with the tokens after the edit location
instead.

\MyPara{Replacement}
If the span of tokens to be replaced is unique in the old sequence,
regular \CodeIn{Replace} sequence is sufficient and deterministic; in
that case we will keep using it.  Otherwise, it is unclear which
occurrence of token span should be replaced.  As an example, in
Figure~\ref{fig:intro-example}, the \Java{} code change is changing
from \CodeIn{PdfException} to \CodeIn{LayoutExceptionMessageConstant}
in the \CodeIn{assertEquals} statement on line
\JavaSecondReplaceLineNumber.
The replacement in the \concise edit sequence is ambiguous because
there are two usages of \CodeIn{PdfException} (on
lines~\JavaFirstReplaceLineNumber{} and \JavaSecondReplaceLineNumber)
in the old \Java{} code sequence after tokenization.
To address this, similar to the insertion case, we search for the
minimal anchor tokens before or after the edit location that can form
a unique span in the old sequence.  For example, the \concise edit sequence:
\MyQuote{
\texttt{<ReplaceOld> PdfException <ReplaceNew>}\\
\texttt{LayoutExceptionMessageConstant <ReplaceEnd>}
}
can be disambiguate into the following \unambig edit sequence:
\MyQuote{
\texttt{<ReplaceOldKeepBefore> format(PdfException}\\
\texttt{<ReplaceNewKeepBefore> format(}\\
\texttt{LayoutExceptionMessageConstant <ReplaceEnd>}
}

\MyPara{Deletion}
Similar to replacement, if the span of tokens to be deleted is unique
across the old sequence, we will keep using \CodeIn{Delete} because it
is unambiguous.  Otherwise, it will be transformed to
\CodeIn{ReplaceKeepBefore} or \CodeIn{ReplaceKeepAfter}.  For example,
suppose the token ``\CodeIn{PdfException.}'' should be removed from
the old \Java{} method on line \JavaSecondReplaceLineNumber{} in
Figure~\ref{fig:intro-example}.  The \concise edit sequence:
\MyQuote{
\texttt{<Delete> PdfException. <DeleteEnd>}
}
will be transformed to:
\MyQuote{
\texttt{<ReplaceOldKeepBefore> format(PdfException.}\\
\texttt{<ReplaceNewKeepBefore> format( <ReplaceEnd>}
} This edit sequence indicates that ``\CodeIn{format(PdfException.}''
should be replaced with ``\CodeIn{format(}'', unambiguously implying
the deletion of ``\CodeIn{PdfException.}''.

To summarize, the \unambig edit sequence contains 4 types of edits:
\texttt{<Replace>}, \texttt{<Delete>}, \texttt{<ReplaceKeepBefore>}
and \texttt{<Re\-place\-KeepAfter>}.  The mappings between \concise
edit sequence and \unambig edit sequence are summarized in
Table~\ref{tab:edit-sequence}.  Given the \unambig edit sequence, we
can apply it to the old input sequence to derive the new edited
sequence deterministically.

\subsection{Model Input}
\label{sec:model-input}

We aim to build performant machine learning models for the \deltatask
task by providing the model with code evolution information, namely
the revisions of code of both source and target programming languages.
Instead of directly translating the entire code snippet between
programming languages, \DeltaTool translates the code \emph{changes}
between programming languages.

\subsubsection{Source Code Edits}
To encourage the model to learn the alignment between developer-made
changes across programming languages, we provide \DeltaTool
with code changes in the source programming language.
To maintain both
precision and conciseness of the edits, we adopt the \unambig edit
sequence (Section~\ref{sec: unambiguous-edits}) to represent the code
changes.  As shown in Figure~\ref{fig:model-architecture}, the \Java{}
code changes (\UseMacro{java_edits}) of replacing the \CodeIn{PdfException} with
\CodeIn{LayoutExceptionMessageConstant} is structured in the form of
\MyQuote{
\UseMacro{intro-example-java-edits}
}

\subsubsection{History-Related Context}

In addition to the learned representation of code changes in source
programming language (\UseMacro{java_edits}), we provide \DeltaTool
with the old code in target programming language (\UseMacro{cs_old})
to better help the model to infer the correlated code changes in the
target programming language.  The intuition is that the model will
reason about how to transfer and tune the edits in source programming
language grounding the specific implementation of the method in target
programming language.

Furthermore, we append the new code in source programming language
(\UseMacro{java_new}) as one of the contexts.  We believe this will
give the model more context to understand the edits in source
programming language and promote the consistency of the updated
methods in two programming languages.

To sum up, we combine history-related context from three sources: code
changes in the source programming language (\UseMacro{java_edits}),
old code in the target programming language (\UseMacro{cs_old}), and
new code in the source programming language (\UseMacro{java_new}).  We
concatenate them into a sequence separated by a special \emph{SEP}
token as the model input.

\subsection{Model Output}
\label{sec:model-output}

We propose two formats as the model's target output which lead to two
\modes of \DeltaTool: \emph{\edittranslation} and \emph{\metaEdits}.
Both \modes use the same input and both \modes' target outputs entail
a sequence of edits on the target programming language.

\MyPara{\edittranslation} The output of \edittranslation \mode is the
\unambig edit sequence in target programming language which suggests
how the code in target programming language should be changed.  Note
that the model-generated \unambig edit sequence can be parsed and
applied to old version of code deterministically.  \edittranslation
essentially learns to translate the code edits from the source
programming language (\UseMacro{java_edits}) to the target programming
language (\UseMacro{cs_edits}) grounding the code history context.
\edittranslation \mode's target output for the \Cs example in
Figure~\ref{fig:intro-example} is:
\MyQuote{
\UseMacro{intro-example-cs-edits}
}

\MyPara{\metaEdits} 
In this \mode, we adopt the output format of
\UseMacro{CoditT5}~\cite{ZhangETAL22CoditT5} for \deltatask since our model is
built upon \UseMacro{CoditT5}, and it had showed promising performance on software editing
tasks.  CoditT5 is \pretrained to generate the following output
format: ``[Edit Plan] <SEP> [Target Sequence]''.  The edit plan is a
\concise edit sequence that represents the steps to edit the input
sequence; the target sequence is the edited sequence after applying
the proceeding edit plan.
We tailored this format to the \deltatask task; the edit plan
represents the edits between the code edits on source programming language
(\UseMacro{java_edits}) and target programming language (\UseMacro{cs_edits}) which we call the \emph{meta edit sequence}.
And the final target sequence should be the \unambig edit sequence on
the target programming language (\UseMacro{cs_edits}).  For the
example in Figure~\ref{fig:intro-example}, the expected meta edit
sequence that converts \Java edit to \Cs edit is the following:
\MyQuote{
\UseMacro{intro-example-meta-edits}
}
The target sequence after applying the meta edit sequence is: 
\MyQuote{
\UseMacro{intro-example-cs-edits}
}
Note that during inference, we only use the target \unambig edit
sequence to get the updated code in target programming language as
\metaEdits mode's prediction.

\section{Dataset}

This is the first work to consider the history of software projects in
a multilingual task; hence, we also created a new dataset that
includes aligned code changes between programming languages.  As the
first step, we build the dataset by mining histories of the
open-source \Java and \Cs projects.   
We first collect the changed methods from the commits of the \Java and
\Cs projects.
We then design heuristics to pair (i.e., align) those changes on
methods with similar implementations and functionalities.
We consider two directions on our dataset:
\UseMacro{java2cs} (updating \Cs method based on \Java changes) and
\UseMacro{cs2java} (updating \Java method based on \Cs changes).  In this section,
we describe the approach we use to collect the data
(Section~\ref{sec:data-collect}), split and preprocess data
(Section~\ref{sec:data-split}), and finally present the statistics of
our dataset (Section~\ref{sec:data-stats}).

\begin{table}[t]
\begin{center}
\caption{\UseMacro{TCap-real-world-dataset-stats}}
\vspace{-8pt}
\begin{small}
\begin{tabular}{l l  r }
\toprule
\textbf{\UseMacro{THead-java-project}}
 & \textbf{\UseMacro{THead-cs-project}}
 & \textbf{\UseMacro{THead-count}}
\\
\midrule
antlr/antlr4
& tunnelvisionlabs/antlr4cs
 & \UseMacro{antlr-antlr4-count}
\\
apache/lucene
& apache/lucenenet
 & \UseMacro{apache-lucene-count}
\\
apache/poi
&nissl-lab/npoi
 & \UseMacro{apache-poi-count}
\\
eclipse/jgit
& mono/ngit
 & \UseMacro{eclipse-jgit-count}
\\
formicary/fpml-toolkit-java
& formicary/fpml-toolkit-csharp
 & \UseMacro{formicary-fpml-toolkit-java-count}
\\
itext/itext7
& itext/itext7-dotnet
 & \UseMacro{itext-itext7-count}
 \\
quartz-scheduler/quartz
& quartznet/quartznet
 & \UseMacro{quartz-scheduler-quartz-count}
\\
terabyte/jgit
& mono/ngit
 & \UseMacro{terabyte-jgit-count}
\\
\midrule
\textbf{SUM}
&
& \Datasize\\
\bottomrule
\end{tabular}
\vspace{\UseMacro{TVspace-real-world-dataset-stats}}
\end{small}
\end{center}
\end{table}

\begin{table}[t]
\begin{center}
\caption{\UseMacro{TCap-dataset-edits-stats}}
\vspace{-8pt}
\begin{tabular}{l  l | r  r  r }
\toprule
 & 
 & \textbf{\UseMacro{THead-train}}
 & \textbf{\UseMacro{THead-valid}}
 & \textbf{\UseMacro{THead-test}}
\\
\midrule
\multicolumn{2}{l|}{\UseMacro{THead-count}}
 & \UseMacro{ds-train-count}
 & \UseMacro{ds-valid-count}
 & \UseMacro{ds-test-count}
 \\
 \midrule
\multirow{5}{*}{Java}
& \UseMacro{THead-mean-old-method-size}
 & \UseMacro{train-java-mean-old-method-size}
 & \UseMacro{valid-java-mean-old-method-size}
 & \UseMacro{test-java-mean-old-method-size}
\\
&\UseMacro{THead-mean-new-method-size}
 & \UseMacro{train-java-mean-new-method-size}
 & \UseMacro{valid-java-mean-new-method-size}
 & \UseMacro{test-java-mean-new-method-size}
\\
&\UseMacro{THead-mean-edit-actions}
 & \UseMacro{train-java-mean-edit-actions}
 & \UseMacro{valid-java-mean-edit-actions}
 & \UseMacro{test-java-mean-edit-actions}
\\
&\UseMacro{THead-mean-add-tokens}
 & \UseMacro{train-java-mean-add-tokens}
 & \UseMacro{valid-java-mean-add-tokens}
 & \UseMacro{test-java-mean-add-tokens}
\\
&\UseMacro{THead-mean-del-tokens}
 & \UseMacro{train-java-mean-del-tokens}
 & \UseMacro{valid-java-mean-del-tokens}
 & \UseMacro{test-java-mean-del-tokens}
\\
\midrule
\multirow{5}{*}{C\#}
&\UseMacro{THead-mean-old-method-size}
 & \UseMacro{train-cs-mean-old-method-size}
 & \UseMacro{valid-cs-mean-old-method-size}
 & \UseMacro{test-cs-mean-old-method-size}
\\
&\UseMacro{THead-mean-new-method-size}
 & \UseMacro{train-cs-mean-new-method-size}
 & \UseMacro{valid-cs-mean-new-method-size}
 & \UseMacro{test-cs-mean-new-method-size}
\\
&\UseMacro{THead-mean-edit-actions}
 & \UseMacro{train-cs-mean-edit-actions}
 & \UseMacro{valid-cs-mean-edit-actions}
 & \UseMacro{test-cs-mean-edit-actions}
\\
&\UseMacro{THead-mean-add-tokens}
 & \UseMacro{train-cs-mean-add-tokens}
 & \UseMacro{valid-cs-mean-add-tokens}
 & \UseMacro{test-cs-mean-add-tokens}
\\
&\UseMacro{THead-mean-del-tokens}
 & \UseMacro{train-cs-mean-del-tokens}
 & \UseMacro{valid-cs-mean-del-tokens}
 & \UseMacro{test-cs-mean-del-tokens}
\\
\bottomrule
\end{tabular}
\vspace{\UseMacro{TVspace-dataset-edits-stats}}
\end{center}
\end{table}

\subsection{Data Collection}
\label{sec:data-collect}
To build the dataset, we extract aligned \Java and \Cs code changes at
the method level as tuples (\Java old method; \Java new method, \Cs
old method; \Cs new method).
The code changes are mined from the git commits.  We consider
\ProjectsNum{} open-source projects as listed in
Table~\ref{tab:real-world-dataset-stats} which have both \Java and \Cs
implementations and are used in prior work~\cite{NguyenETAL15Divide,
  CodeXGLUE, ChenETAL18Tree}.  All the projects were first developed
in \Java and then ported to \Cs.

To collect the paired changes, we first assign a unique identifier to
each method in the projects (for both \Java and \Cs projects) based on
the method signature, class name and path to the file where the method
is defined.  Similar to the strategy used by \citet{CodeXGLUE}, we
then pair the \Java methods and \Cs methods according to the
similarity of their unique identifiers.  This strategy is effective
because the ported \Cs project has very similar structure and naming
rules for classes and methods to the corresponding \Java project.

We use the following rules to extract the aligned code changes:
\begin{enumerate}[topsep=3pt,itemsep=1ex,partopsep=0ex,parsep=0ex,leftmargin=*]
\item For each \Java method change, we extract the code changes in the
paired \Cs method that happen no later than 90 days of the \Java
change as the \emph{possible matched code change}. We use the commit
date as the time of the change.
\item To filter unrelated code changes, we compute the Jaccard similarity~\cite{Jaccard} between \Cs and \Java added and deleted lines.
We further refine the filtering by sub-tokenizing these lines based on camelCase conventions (e.g., \CodeIn{lastModified} to
\CodeIn{last}\ \CodeIn{modi}\-\CodeIn{fied}) and compute Jaccard similarity only for the added and deleted tokens.
We only keep possible matched code changes that have the token-level Jaccard similarity higher than \tokenJaccardThreshold{} and the line-level Jaccard similarity higher than \lineJaccardThreshold{}.
\item For each \Java code change and \Cs code change, we only select
the most similar corresponding code change if there are multiple
possible matched code changes.
\end{enumerate}

\subsection{Data Preprocessing and Splitting}
\label{sec:data-split}

For both \Java and \Cs methods, we remove the inline natural language
comments and tokenize the method into tokens using the
language-specific lexers generated by Antlr~\cite{antlr}.

We envision the following use case for the machine learning model:
whenever a developer makes a change in the project written in the source
programming language, the developer will use the model trained on the
existing historical aligned code changes to migrate that change to
projects written in other target programming languages.  To evaluate
the models under this use case, following the recommendations from prior
work~\cite{NieETAL22EvalMethodologies}, we split the dataset into
training, validation and test sets using the \emph{time-segmented} approach.
Namely, the changes in the training set took place before the changes
in the validation set, which in turn took place before the changes in
the test set.  More specific, for each \Java and \Cs code change
pair, we first collect the time of the \Cs commit and then sort the
code change pairs in chronological order.  We then select the oldest
\TrainRatio{} of the code change pairs from each project as training
data, next oldest \ValRatio{} as validation data, the remaining as test
data.

To more rigorously assess the generalization capabilities of the
  models, we also evaluated them when splitting the dataset using the
  \emph{cross-project} approach~\cite{NieETAL22EvalMethodologies},
  which is frequently used in prior work on machine learning models
  for code.
Specifically, the aligned code changes in the training set
are from different projects compared to those in the validation and test sets.

\subsection{Statistics}
\label{sec:data-stats}
The statistics of the collected dataset are shown in
Table~\ref{tab:dataset-edits-stats}.  We present the number of examples in the training,
validation, and test dataset using time-segmented split approach.
We show the average number of tokens in the old methods (Avg.
len(\UseMacro{code_old})) and new methods (Avg.
len(\UseMacro{code_new})) after tokenization by the lexers.  To
measure the size of the code changes, we calculate the average number
of added tokens (Avg. \# add. tks) and deleted tokens (Avg. \# del.
tks) in the changed \Java and \Cs methods as well as the average
number of edits (Avg. \# edits) needed for those changes.  
For computing these edit-related statistics, we represent the code
changes using \concise edit sequences
(Section~\ref{sec:edits-sequence}).

For both \Java and \Cs code changes, the difference between average
number of added tokens and deleted tokens is usually
small, fewer than \TokenDiff{} tokens.  Similarly, we find that the
average number of edits needed is fewer than \EditAct{} and the edits
happened in the newer commits are generally smaller
than prior ones.  This is expected as the software projects are
becoming more stable as they evolve, and thus there will be smaller code
changes to be made.  For evaluation, we run all the models and
baselines on this dataset in two directions: (1)~updating \Cs method based on \Java changes,
 and (2)~updating \Java method based on \Cs changes.  We denote the former
one as \UseMacro{java2cs} and the latter one as \UseMacro{cs2java}.

\section{Experiments}
In this section, we describe the baselines we compare to with our
\DeltaTool model (Section~\ref{sec: baselines}), the evaluation
metrics (Section~\ref{sec: metrics}) and the detailed experiment setup
(Section~\ref{sec: setup}).

\subsection{Baselines}
\label{sec: baselines}
We evaluate our approach against rule-based models, \pretrained
encoder-decoder models, the state-of-the-art code-editing model (which
targets a single programming language), and large generative models \pretrained on
billions of lines of code.

\MyPara{\Copy}
This is a rule-based model which copies the old code in target
programming language (\UseMacro{cs_old}) as the prediction.  This is
not a trivial baseline since there are quite a few examples in the
dataset that entail small edits between two versions.  We include this
to benchmark the models that actually update the code.

\MyPara{\CopyEdits}
Based on our observations, there are cases where the code change in
source programming language (\UseMacro{java_edits}) is exactly the
same as the change in target programming language
(\UseMacro{cs_edits}) , such as changing the variable name or updating
the log message.  This rule-based model copies the
\UseMacro{java_edits} and directly applies it to the old code in target
programming language (\UseMacro{cs_old}).

\MyPara{\translationModel}
We consider a state-of-the-art model that does not have access to the
code change history. Namely, a code translation model that translates
code between the programming languages (from \UseMacro{java_new} to
\UseMacro{cs_new}).  We use CodeT5~\cite{WangETAL21codet5}, an \LLM
\pretrained on large amount of developer-written code from GitHub,
which we \finetune on our constructed dataset.

\MyPara{\UseMacro{CodeT5Up}}
This model has the same architecture as \translationModel except that
we supply it with code change history.  The model input is the same as
for our \DeltaTool models, i.e., with extra context of the old code in
target programming language (\UseMacro{cs_old}) and the code change in
source programming language (\UseMacro{java_edits}).  Different from
\DeltaTool model, it is trained to directly generate the new code in
target programming language (\UseMacro{cs_new}).

\MyPara{\UseMacro{CoditT5}}
This is the state-of-the-art model for software editing
tasks~\cite{ZhangETAL22CoditT5}.  It has the same model architecture and input
as \DeltaTool, while the output consists of the
edit plan to represent the edits on the target programming language
and the target sequence which represents the updated code
(\UseMacro{cs_new}) after applying the edit plan.

\MyPara{\Codex-few-shot~\cite{Codex}}
Large \pretrained generative models such as GPT-3~\cite{GPT3} have
shown impressive results under the context of \emph{few-shot learning}
or even \emph{zero-shot learning} on various generation tasks.
They are able to generalize to new tasks they have not seen during
\pretraining with only a few or even no labeled examples.  To compare
the \finetuned \DeltaTool model with generative models, we include
\Codex, a large generative model built on GPT-3 and is
further \pretrained on billions of GitHub data.  Following prior
work~\cite{Ahmed22Few,Khan22Automatic}, for each example in test data,
we randomly select several labeled examples in the training data as
the context.  Note that the labeled examples are selected from the
same project as the test data.  For \UseMacro{java2cs} dataset, each
labeled example is formed as: ``\Java: \UseMacro{java_old} =>
\UseMacro{java_new} \Cs: \UseMacro{cs_old} => \UseMacro{cs_new}'' to
inform the model the aligned updates between two programming
languages.  The designed prompt for inference is ``\Java:
\UseMacro{java_old} => \UseMacro{java_new} \Cs: \UseMacro{cs_old} =>
''.  The model output is the prediction for the new code in target
programming language (\UseMacro{cs_new}).
To conform to the required input length limit, we include 2 labeled
examples in the prompt.

\MyPara{\chatgpt-zero-shot~\cite{OpenAIChatGPT}}
\chatgpt is an upgraded version of GPT-3 model and is further fine-tuned
for conversation generation following human instructions with the help of supervised and reinforcement learning 
methods.
It has showed strong performance on code completion benchmarks like HumanEval and MBPP~\cite{Codex, MBPP,GPT4Report}.
For each example in test data, we provide instructions including both the previous and the updated versions of the code written in the source programming language, subsequently prompting \chatgpt to update the old code in the target programming language accordingly.
For \UseMacro{java2cs} dataset, the prompt is formed as: ``The developer updates the Java method from: \UseMacro{java_old} to:  \UseMacro{java_new}. Please update the C\# method accordingly. This is the old C\# method: \UseMacro{cs_old}.''
\vspace{-5pt}
\subsection{Evaluation Metrics}
\label{sec: metrics}
Following prior work~\cite{WangETAL21codet5, ZhangETAL22CoditT5,
PanthaplackelETAL20Learning, PLBART}, we use metrics for evaluating the
quality of code generation: BLEU~\cite{papineni2002bleu},
CodeBLEU~\cite{CodeBLEU}, xMatch, and metrics for evaluating the
quality of software editing: SARI~\cite{xu2016optimizing} and
GLEU~\cite{napoles2015ground}.  Note that for all the metrics we
report in this paper, they range from 0 to 100 and higher scores are
better.

\MyPara{xMatch}
When the generated code matches exactly with the expected code in
target programming language, this metric is 100; otherwise, this
metric is 0. This metric reflects the percentage of exact matches
among the models' predictions on test data.

\MyPara{BLEU}
It is a widely used metric originally proposed for evaluating the
quality of machine translation.  It measures the n-gram overlap
between the generated sequence and the expected one.  Concretely, we
report the 1$\sim$4-grams overlap between the tokens in the
predictions and tokens in the ground truth.

\MyPara{CodeBLEU}
The metric is proposed for evaluating the quality of code generation.
In addition to measuring the n-gram overlap, it considers the overlap
of the Abstract Syntax Tree (AST) and data-flow graph between
generated code and the expected code.

\MyPara{SARI}
It measures quality of the systems that are designed to make edits.
Specifically, it is computed as the average of the F1 score for kept
and inserted spans of tokens, and the precision of deleted spans of
tokens.

\MyPara{GLEU}
It is a variant of BLEU.  It was originally proposed for grammatical
error correction and designed for rewarding the correct edits while
penalizing the incorrect ones.

\subsection{Experimental Setup}
\label{sec: setup}
We run all experiments on machines with 4 NVidia 1080-TI GPUs,
Intel(R) Xeon(R) CPU E5-2620 v4 @ 2.10GHz for training.  We implement
our models using PyTorch 1.9.0.  
All the hyper-parameters of the
CodeT5 and CoditT5 baselines are set to the same values as in prior
work~\cite{WangETAL21codet5, ZhangETAL22CoditT5}.
For \DeltaTool, \translationModel, \UseMacro{CodeT5Up}, and \UseMacro{CoditT5}, we early
stop the training when the BLEU score on the validation set does not improve for 5
epochs, and use beam search with beam size 20 during inference.
For \Codex and \chatgpt, we set temperature to 0.2 during inference.

Note that \Codex and \chatgpt are closed-source and may be
  updated/deprecated over time.  We used the code-davinci-002 version
  of \Codex when performing experiments in the time-segmented split;
  however, OpenAI deprecated \Codex in March 2023 before we could
  complete our experiments in the cross-project split, as such we did
  not include \Codex in this part of results.

\section{Results}
\label{sec:eval}

We organize our evaluation around three main research questions:

\RQ{1}{What is the benefit of using code change history in
\deltatask?}

\RQ{2}{How does our edit-based model, \DeltaTool, compare to
  generation-based models for the \deltatask?}

\RQ{3}{How can a generation-based model complement \DeltaTool model to
  further improve the performance?}

\subsection{Quantitative Analysis}
\label{sec:eval:quantitative}

In
tables~\ref{tab:java2cstranslation-results}-\ref{tab:cs2java-cp-translation-results},
we present results for baselines and our proposed \DeltaTool models on
\UseMacro{java2cs}, \UseMacro{cs2java} for both time-segmented and
cross-project splits.
We conducted statistical significance testing through bootstrap
tests~\cite{Berg-KirkpatrickETAL12Empirical} under confidence level
95\%.

\begin{table*}[t]
\begin{center}
\caption{\UseMacro{TCap-java2cs-translation-models-results}}
\vspace{-8pt}
\begin{tabular}{l | c  c  c  c  c  c }
\toprule
\textbf{\UseMacro{THead-models}}
 & \textbf{\UseMacro{THead-Acc-top-1}}
 & \textbf{\UseMacro{THead-BLEU}}
 & \textbf{\UseMacro{THead-CodeBLEU}}
 & \textbf{\UseMacro{THead-SARI}}
 & \textbf{\UseMacro{THead-GLEU}}
\\
\midrule
\UseMacro{THead-copy}
 & \UseMacro{res-java2cs-copy-Acc-top-1}
 & \UseMacro{res-java2cs-copy-BLEU}
 & \UseMacro{res-java2cs-copy-CodeBLEU}
 & \UseMacro{res-java2cs-copy-SARI}
 & \UseMacro{res-java2cs-copy-GLEU}
\\
\UseMacro{THead-CopyEdits}
 & \UseMacro{res-java2cs-CopyEdits-Acc-top-1}$^{\beta}$
 & \UseMacro{res-java2cs-CopyEdits-BLEU}$^{\alpha}$$^{\chi}$
 & \UseMacro{res-java2cs-CopyEdits-CodeBLEU}
 & \UseMacro{res-java2cs-CopyEdits-SARI}
 & \UseMacro{res-java2cs-CopyEdits-GLEU}
\\
\midrule
\UseMacro{THead-translation-CodeT5}
 & \UseMacro{res-java2cs-translation-CodeT5-Acc-top-1}$^{\beta}$
 & \UseMacro{res-java2cs-translation-CodeT5-BLEU}
 & \UseMacro{res-java2cs-translation-CodeT5-CodeBLEU}
 & \UseMacro{res-java2cs-translation-CodeT5-SARI}
 & \UseMacro{res-java2cs-translation-CodeT5-GLEU}
\\
\UseMacro{THead-update-CodeT5}
 & \UseMacro{res-java2cs-update-CodeT5-Acc-top-1}$^{\epsilon}$
 & \UseMacro{res-java2cs-update-CodeT5-BLEU}$^{\chi}$$^{\eta}$
 & \UseMacro{res-java2cs-update-CodeT5-CodeBLEU}
 & \UseMacro{res-java2cs-update-CodeT5-SARI}
 & \UseMacro{res-java2cs-update-CodeT5-GLEU}
\\
\UseMacro{THead-CoditT5}
 & \UseMacro{res-java2cs-CoditT5-Acc-top-1}$^{\epsilon}$
 & \UseMacro{res-java2cs-CoditT5-BLEU}$^{\alpha}$$^{\eta}$
 & \UseMacro{res-java2cs-CoditT5-CodeBLEU}
 & \UseMacro{res-java2cs-CoditT5-SARI}
 & \UseMacro{res-java2cs-CoditT5-GLEU}
\\
\UseMacro{THead-codex}
 & \UseMacro{res-java2cs-codex-Acc-top-1}
 & \UseMacro{res-java2cs-codex-BLEU}
 & \UseMacro{res-java2cs-codex-CodeBLEU}
 & \UseMacro{res-java2cs-codex-SARI}
 & \UseMacro{res-java2cs-codex-GLEU}
\\
\UseMacro{THead-chatgpt3.5-zero-shot}
 & \UseMacro{res-java2cs-chatgpt3.5-zero-shot-Acc-top-1}
 & \UseMacro{res-java2cs-chatgpt3.5-zero-shot-BLEU}
 & \UseMacro{res-java2cs-chatgpt3.5-zero-shot-CodeBLEU}
 & \UseMacro{res-java2cs-chatgpt3.5-zero-shot-SARI}
 & \UseMacro{res-java2cs-chatgpt3.5-zero-shot-GLEU}
\\
\midrule
\UseMacro{THead-metaEdits}
 & \UseMacro{res-java2cs-metaEdits-Acc-top-1}
 & \UseMacro{res-java2cs-metaEdits-BLEU}
 & \UseMacro{res-java2cs-metaEdits-CodeBLEU}
 & \UseMacro{res-java2cs-metaEdits-SARI}
 & \UseMacro{res-java2cs-metaEdits-GLEU}
\\
\UseMacro{THead-edit-translation}
 & \UseMacro{res-java2cs-edit-translation-Acc-top-1}
 & \UseMacro{res-java2cs-edit-translation-BLEU}
 & \UseMacro{res-java2cs-edit-translation-CodeBLEU}
 & \textbf{\UseMacro{res-java2cs-edit-translation-SARI}}$^{\delta}$
 & \UseMacro{res-java2cs-edit-translation-GLEU}
\\
\midrule
\UseMacro{THead-hybrid}
 & \textbf{\UseMacro{res-java2cs-hybrid-Acc-top-1}}
 & \textbf{\UseMacro{res-java2cs-hybrid-BLEU}}
 & \textbf{\UseMacro{res-java2cs-hybrid-CodeBLEU}}
 & \UseMacro{res-java2cs-hybrid-SARI}$^{\delta}$
 & \textbf{\UseMacro{res-java2cs-hybrid-GLEU}}
\\
\bottomrule
\end{tabular}
\label{tab:java2cstranslation-results}
\end{center}
\end{table*}

\begin{table*}[t]
\begin{center}
\caption{\UseMacro{TCap-cs2java-translation-models-results}}
\vspace{-8pt}
\begin{tabular}{l | c  c  c  c  c  c }
\toprule
\textbf{\UseMacro{THead-models}}
 & \textbf{\UseMacro{THead-Acc-top-1}}
 & \textbf{\UseMacro{THead-BLEU}}
 & \textbf{\UseMacro{THead-CodeBLEU}}
 & \textbf{\UseMacro{THead-SARI}}
 & \textbf{\UseMacro{THead-GLEU}}
\\
\midrule
\UseMacro{THead-copy}
 & \UseMacro{res-cs2java-copy-Acc-top-1}
 & \UseMacro{res-cs2java-copy-BLEU}
 & \UseMacro{res-cs2java-copy-CodeBLEU}
 & \UseMacro{res-cs2java-copy-SARI}
 & \UseMacro{res-cs2java-copy-GLEU}
\\
\UseMacro{THead-CopyEdits}
 & \UseMacro{res-cs2java-CopyEdits-Acc-top-1}
 & \UseMacro{res-cs2java-CopyEdits-BLEU}$^{\alpha}$
 & \UseMacro{res-cs2java-CopyEdits-CodeBLEU}$^{\beta}$
 & \UseMacro{res-cs2java-CopyEdits-SARI}
 & \UseMacro{res-cs2java-CopyEdits-GLEU}$^{\chi}$
\\
\midrule
\UseMacro{THead-translation-CodeT5}
 & \UseMacro{res-cs2java-translation-CodeT5-Acc-top-1}
 & \UseMacro{res-cs2java-translation-CodeT5-BLEU}$^{\alpha}$
 & \UseMacro{res-cs2java-translation-CodeT5-CodeBLEU}$^{\beta}$
 & \UseMacro{res-cs2java-translation-CodeT5-SARI}
 & \UseMacro{res-cs2java-translation-CodeT5-GLEU}$^{\chi}$
\\
\UseMacro{THead-update-CodeT5}
 & \UseMacro{res-cs2java-update-CodeT5-Acc-top-1}
 & \UseMacro{res-cs2java-update-CodeT5-BLEU}
 & \UseMacro{res-cs2java-update-CodeT5-CodeBLEU}$^{\gamma}$
 & \UseMacro{res-cs2java-update-CodeT5-SARI}
 & \UseMacro{res-cs2java-update-CodeT5-GLEU}
\\
\UseMacro{THead-CoditT5}
 & \UseMacro{res-cs2java-CoditT5-Acc-top-1}
 & \UseMacro{res-cs2java-CoditT5-BLEU}
 & \UseMacro{res-cs2java-CoditT5-CodeBLEU}$^{\gamma}$
 & \UseMacro{res-cs2java-CoditT5-SARI}
 & \UseMacro{res-cs2java-CoditT5-GLEU}
\\
\UseMacro{THead-codex}
 & \UseMacro{res-cs2java-codex-Acc-top-1}
 & \UseMacro{res-cs2java-codex-BLEU}
 & \UseMacro{res-cs2java-codex-CodeBLEU}
 & \UseMacro{res-cs2java-codex-SARI}
 & \UseMacro{res-cs2java-codex-GLEU}
\\
\UseMacro{THead-chatgpt3.5-zero-shot}
 & \UseMacro{res-cs2java-chatgpt3.5-zero-shot-Acc-top-1}
 & \UseMacro{res-cs2java-chatgpt3.5-zero-shot-BLEU}
 & \UseMacro{res-cs2java-chatgpt3.5-zero-shot-CodeBLEU}
 & \UseMacro{res-cs2java-chatgpt3.5-zero-shot-SARI}
 & \UseMacro{res-cs2java-chatgpt3.5-zero-shot-GLEU}
\\
\midrule
\UseMacro{THead-metaEdits}
 & \textbf{\UseMacro{res-cs2java-metaEdits-Acc-top-1}}$^{\epsilon}$$^{\eta}$
 & \UseMacro{res-cs2java-metaEdits-BLEU}
 & \UseMacro{res-cs2java-metaEdits-CodeBLEU}
 & \UseMacro{res-cs2java-metaEdits-SARI}
 & \UseMacro{res-cs2java-metaEdits-GLEU}
\\
\UseMacro{THead-edit-translation}
 & \UseMacro{res-cs2java-edit-translation-Acc-top-1}$^{\delta}$$^{\epsilon}$
 & \UseMacro{res-cs2java-edit-translation-BLEU}
 & \UseMacro{res-cs2java-edit-translation-CodeBLEU}
 & \textbf{\UseMacro{res-cs2java-edit-translation-SARI}}
 & \UseMacro{res-cs2java-edit-translation-GLEU}
\\
\midrule
\UseMacro{THead-hybrid}
 & \UseMacro{res-cs2java-hybrid-Acc-top-1}$^{\delta}$$^{\eta}$
 & \textbf{\UseMacro{res-cs2java-hybrid-BLEU}}
 & \textbf{\UseMacro{res-cs2java-hybrid-CodeBLEU}}
 & \UseMacro{res-cs2java-hybrid-SARI}
 & \textbf{\UseMacro{res-cs2java-hybrid-GLEU}}
\\
\bottomrule
\end{tabular}
\vspace{\UseMacro{TVspace-cs2java-translation-models-results}}
\end{center}
\end{table*}

\begin{table*}[t]
\begin{center}
\caption{\UseMacro{TCap-java2cs-cp-translation-models-results}}
\vspace{-8pt}
\begin{tabular}{l | c  c  c  c  c  c }
\toprule
\textbf{\UseMacro{THead-models}}
 & \textbf{\UseMacro{THead-Acc-top-1}}
 & \textbf{\UseMacro{THead-BLEU}}
 & \textbf{\UseMacro{THead-CodeBLEU}}
 & \textbf{\UseMacro{THead-SARI}}
 & \textbf{\UseMacro{THead-GLEU}}
\\
\midrule
\UseMacro{THead-copy}
 & \UseMacro{res-java2cs-cp-copy-Acc-top-1}
 & \UseMacro{res-java2cs-cp-copy-BLEU}$^{\alpha}$
 & \UseMacro{res-java2cs-cp-copy-CodeBLEU}
 & \UseMacro{res-java2cs-cp-copy-SARI}
 & \UseMacro{res-java2cs-cp-copy-GLEU}
\\
\UseMacro{THead-CopyEdits}
 & \UseMacro{res-java2cs-cp-CopyEdits-Acc-top-1}
 & \UseMacro{res-java2cs-cp-CopyEdits-BLEU}
 & \UseMacro{res-java2cs-cp-CopyEdits-CodeBLEU}
 & \UseMacro{res-java2cs-cp-CopyEdits-SARI}
 & \UseMacro{res-java2cs-cp-CopyEdits-GLEU}
\\
\midrule
\UseMacro{THead-translation-CodeT5}
 & \UseMacro{res-java2cs-cp-translation-CodeT5-Acc-top-1}
 & \UseMacro{res-java2cs-cp-translation-CodeT5-BLEU}
 & \UseMacro{res-java2cs-cp-translation-CodeT5-CodeBLEU}
 & \UseMacro{res-java2cs-cp-translation-CodeT5-SARI}
 & \UseMacro{res-java2cs-cp-translation-CodeT5-GLEU}
\\
\UseMacro{THead-update-CodeT5}
 & \UseMacro{res-java2cs-cp-update-CodeT5-Acc-top-1}
 & \UseMacro{res-java2cs-cp-update-CodeT5-BLEU}$^{\alpha}$
 & \UseMacro{res-java2cs-cp-update-CodeT5-CodeBLEU}
 & \UseMacro{res-java2cs-cp-update-CodeT5-SARI}
 & \UseMacro{res-java2cs-cp-update-CodeT5-GLEU}
\\
\UseMacro{THead-CoditT5}
 & \UseMacro{res-java2cs-cp-CoditT5-Acc-top-1}
 & \UseMacro{res-java2cs-cp-CoditT5-BLEU}
 & \UseMacro{res-java2cs-cp-CoditT5-CodeBLEU}
 & \UseMacro{res-java2cs-cp-CoditT5-SARI}
 & \UseMacro{res-java2cs-cp-CoditT5-GLEU}
\\
\UseMacro{THead-chatgpt3.5-zero-shot}
 & \UseMacro{res-java2cs-cp-chatgpt3.5-zero-shot-Acc-top-1}
 & \UseMacro{res-java2cs-cp-chatgpt3.5-zero-shot-BLEU}
 & \UseMacro{res-java2cs-cp-chatgpt3.5-zero-shot-CodeBLEU}
 & \UseMacro{res-java2cs-cp-chatgpt3.5-zero-shot-SARI}
 & \UseMacro{res-java2cs-cp-chatgpt3.5-zero-shot-GLEU}
\\
\midrule
\UseMacro{THead-metaEdits}
 & \UseMacro{res-java2cs-cp-metaEdits-Acc-top-1}
 & \UseMacro{res-java2cs-cp-metaEdits-BLEU}$^{\beta}$
 & \UseMacro{res-java2cs-cp-metaEdits-CodeBLEU}
 & \UseMacro{res-java2cs-cp-metaEdits-SARI}
 & \UseMacro{res-java2cs-cp-metaEdits-GLEU}$^{\chi}$
\\
\UseMacro{THead-edit-translation}
 & \UseMacro{res-java2cs-cp-edit-translation-Acc-top-1}
 & \UseMacro{res-java2cs-cp-edit-translation-BLEU}$^{\beta}$
 & \UseMacro{res-java2cs-cp-edit-translation-CodeBLEU}
 & \UseMacro{res-java2cs-cp-edit-translation-SARI}
 & \UseMacro{res-java2cs-cp-edit-translation-GLEU}$^{\chi}$
\\
\midrule
\UseMacro{THead-hybrid}
 & \textbf{\UseMacro{res-java2cs-cp-hybrid-Acc-top-1}}
 & \textbf{\UseMacro{res-java2cs-cp-hybrid-BLEU}}
 & \textbf{\UseMacro{res-java2cs-cp-hybrid-CodeBLEU}}
 & \textbf{\UseMacro{res-java2cs-cp-hybrid-SARI}}
 & \textbf{\UseMacro{res-java2cs-cp-hybrid-GLEU}}
\\
\bottomrule
\end{tabular}
\label{tab:java2cs-cp-translation-results}
\end{center}
\end{table*}

\begin{table*}[t]
\begin{center}
\caption{\UseMacro{TCap-cs2java-cp-translation-models-results}}
\vspace{-8pt}
\begin{tabular}{l | c  c  c  c  c  c }
\toprule
\textbf{\UseMacro{THead-models}}
 & \textbf{\UseMacro{THead-Acc-top-1}}
 & \textbf{\UseMacro{THead-BLEU}}
 & \textbf{\UseMacro{THead-CodeBLEU}}
 & \textbf{\UseMacro{THead-SARI}}
 & \textbf{\UseMacro{THead-GLEU}}
\\
\midrule
\UseMacro{THead-copy}
 & \UseMacro{res-cs2java-cp-copy-Acc-top-1}
 & \UseMacro{res-cs2java-cp-copy-BLEU}
 & \UseMacro{res-cs2java-cp-copy-CodeBLEU}
 & \UseMacro{res-cs2java-cp-copy-SARI}
 & \UseMacro{res-cs2java-cp-copy-GLEU}
\\
\UseMacro{THead-CopyEdits}
 & \UseMacro{res-cs2java-cp-CopyEdits-Acc-top-1}
 & \UseMacro{res-cs2java-cp-CopyEdits-BLEU}
 & \UseMacro{res-cs2java-cp-CopyEdits-CodeBLEU}
 & \UseMacro{res-cs2java-cp-CopyEdits-SARI}
 & \UseMacro{res-cs2java-cp-CopyEdits-GLEU}
\\
\midrule
\UseMacro{THead-translation-CodeT5}
 & \UseMacro{res-cs2java-cp-translation-CodeT5-Acc-top-1}
 & \UseMacro{res-cs2java-cp-translation-CodeT5-BLEU}
 & \UseMacro{res-cs2java-cp-translation-CodeT5-CodeBLEU}
 & \UseMacro{res-cs2java-cp-translation-CodeT5-SARI}
 & \UseMacro{res-cs2java-cp-translation-CodeT5-GLEU}
\\
\UseMacro{THead-update-CodeT5}
 & \UseMacro{res-cs2java-cp-update-CodeT5-Acc-top-1}
 & \UseMacro{res-cs2java-cp-update-CodeT5-BLEU}
 & \UseMacro{res-cs2java-cp-update-CodeT5-CodeBLEU}
 & \UseMacro{res-cs2java-cp-update-CodeT5-SARI}
 & \UseMacro{res-cs2java-cp-update-CodeT5-GLEU}
\\
\UseMacro{THead-CoditT5}
 & \UseMacro{res-cs2java-cp-CoditT5-Acc-top-1}
 & \UseMacro{res-cs2java-cp-CoditT5-BLEU}
 & \UseMacro{res-cs2java-cp-CoditT5-CodeBLEU}
 & \UseMacro{res-cs2java-cp-CoditT5-SARI}
 & \UseMacro{res-cs2java-cp-CoditT5-GLEU}
\\
\UseMacro{THead-chatgpt3.5-zero-shot}
 & \UseMacro{res-cs2java-cp-chatgpt3.5-zero-shot-Acc-top-1}
 & \UseMacro{res-cs2java-cp-chatgpt3.5-zero-shot-BLEU}
 & \UseMacro{res-cs2java-cp-chatgpt3.5-zero-shot-CodeBLEU}
 & \UseMacro{res-cs2java-cp-chatgpt3.5-zero-shot-SARI}
 & \UseMacro{res-cs2java-cp-chatgpt3.5-zero-shot-GLEU}
\\
\midrule
\UseMacro{THead-metaEdits}
 & \UseMacro{res-cs2java-cp-metaEdits-Acc-top-1}
 & \UseMacro{res-cs2java-cp-metaEdits-BLEU}$^{\alpha}$
 & \UseMacro{res-cs2java-cp-metaEdits-CodeBLEU}
 & \UseMacro{res-cs2java-cp-metaEdits-SARI}
 & \UseMacro{res-cs2java-cp-metaEdits-GLEU}$^{\beta}$
\\
\UseMacro{THead-edit-translation}
 & \UseMacro{res-cs2java-cp-edit-translation-Acc-top-1}
 & \UseMacro{res-cs2java-cp-edit-translation-BLEU}$^{\alpha}$
 & \UseMacro{res-cs2java-cp-edit-translation-CodeBLEU}
 & \UseMacro{res-cs2java-cp-edit-translation-SARI}
 & \UseMacro{res-cs2java-cp-edit-translation-GLEU}$^{\beta}$
\\
\midrule
\UseMacro{THead-hybrid}
 & \textbf{\UseMacro{res-cs2java-cp-hybrid-Acc-top-1}}
 & \textbf{\UseMacro{res-cs2java-cp-hybrid-BLEU}}
 & \textbf{\UseMacro{res-cs2java-cp-hybrid-CodeBLEU}}
 & \textbf{\UseMacro{res-cs2java-cp-hybrid-SARI}}
 & \textbf{\UseMacro{res-cs2java-cp-hybrid-GLEU}}
\\
\bottomrule
\end{tabular}
\label{tab:cs2java-cp-translation-results}
\end{center}
\end{table*}

\RQ{1}{\textbf{Conntribution of code change histories.}}
We divide models into two categories with respect to whether a model has
access to the information on code change histories: \Copy and
\translationModel are history-agnostic models, and the remaining are
history-aware models.  Overall, the history-aware models outperform
the history-agnostic ones.  The rule-based model \CopyEdits, which
directly applies the code change in source programming language
(\UseMacro{java_edits}) to the old code in target programming language
without any adaptation, has comparable performance to the
machine learning history-agnostic model \translationModel. This
emphasizes the importance of contextual information provided by code
change histories in \deltatask.
Interestingly, we find that \Codex-few-shot, which is used under the few-shot
learning setting without \finetuning, performs better than \finetuned
\translationModel on xMatch, while worse than other history-aware
\finetuned machine learning models.  This again underlines the value
of code change histories and suggests that \finetuning will give
better performance by leveraging more code history contexts in the
training data.

\begin{figure}[t]
  \centering
    \includegraphics[scale=0.25]{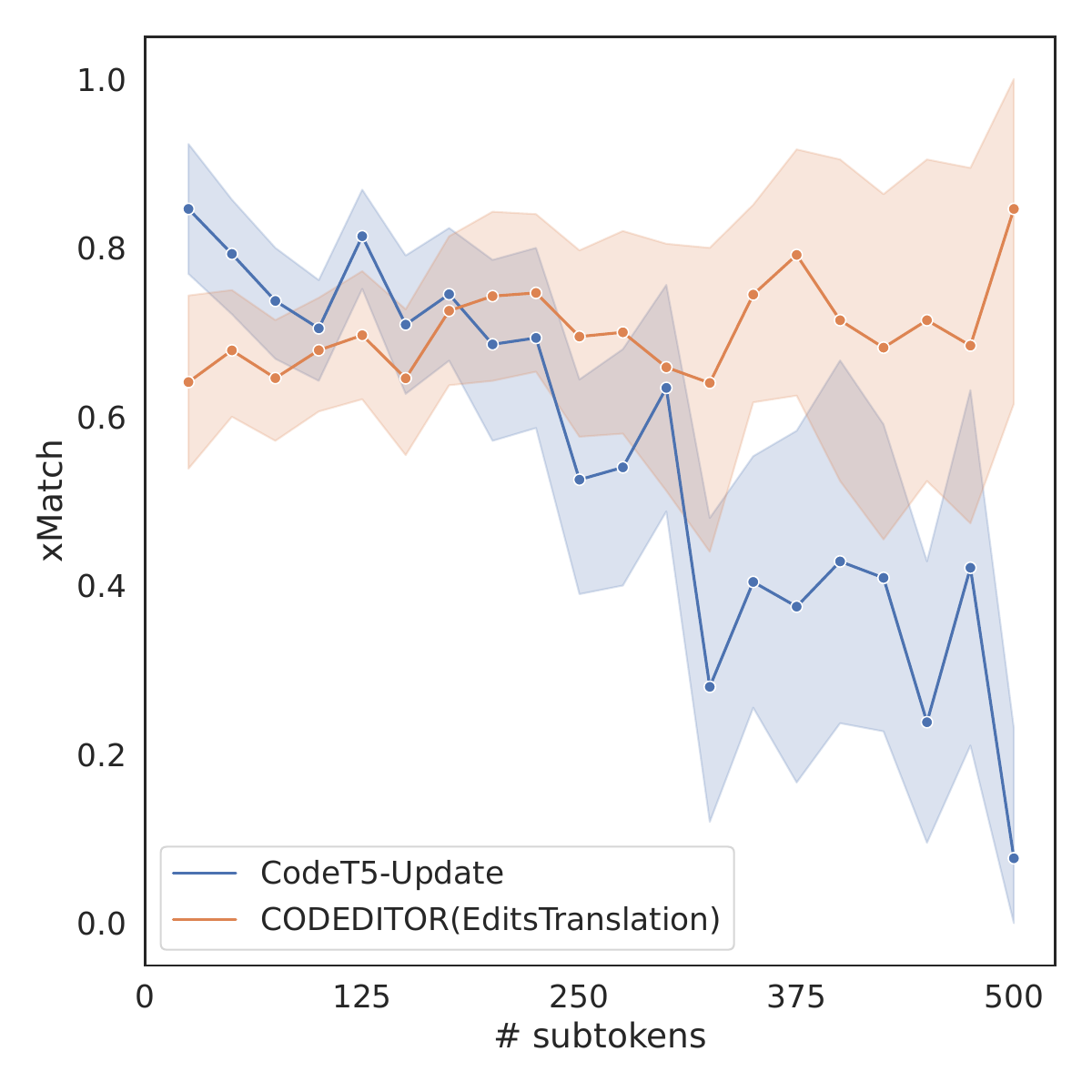}
    \vspace{-12pt}
    \caption{\UseMacro{FCap-xMatch-tokens-number} \label{fig:xmatch-token-number}}
    \vspace{-10pt}
\end{figure}

\begin{figure}[t]
  \centering
    \includegraphics[scale=0.42, trim={0 0 4.5cm 0},clip]{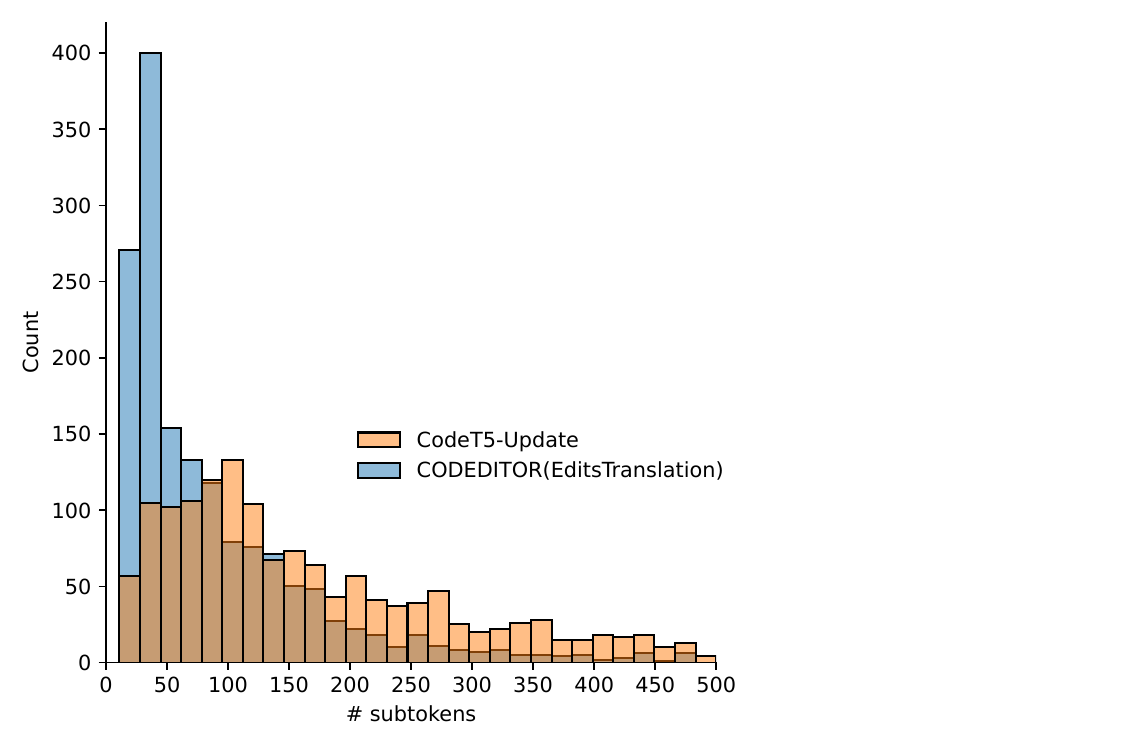}
    \vspace{-12pt}
    \caption{\UseMacro{FCap-ref-tokens-number-dist} \label{fig:ref-token-number}}
    \vspace{-10pt}
\end{figure}

\RQ{2}{\textbf{\DeltaTool vs. generation-based models.}}
Among all the history-aware models, machine learning models, such as
\UseMacro{CodeT5Up} and \UseMacro{CoditT5}, achieve much higher
performance than the rule-based \CopyEdits, which demonstrates that
the machine learning models effectively learns to reason about the
correlated code changes and adjust them to the target programming
language.  We observe that \DeltaTool (in both \edittranslation and
\metaEdits \modes), which is trained to first translate code changes
on source programming language to target programming language and then
apply the edits to the old code in target programming language,
achieve even higher performance across all the metrics than the large
\pretrained generation-based model (\UseMacro{CodeT5Up}) which
directly generates the new code in target programming language from scratch.
This highlights that the models that are trained to explicitly perform
edits by predicting the edit sequence are better suited for editing
tasks in the software domain than generation-based models.

\begin{figure*}[h]
  \centering
  \vspace{5pt}
  \newsavebox\boxJavaCode
\begin{lrbox}{\boxJavaCode}
  \begin{lstlisting}[language=java-pretty, numbers=left]
      public static Document parseBodyFragment(String bodyHtml, String baseUri) {
        ...
        List <Node> nodeList = parseFragment(bodyHtml, body, baseUri);
        Node[] nodes = nodeList.toArray(new Node[0]);
        - for (int i = nodes.length - 1; i > (-w<)nodeList.size()(>w-); i--) {
        + for (int i = nodes.length - 1; i > (+w<)0(>w+); i--) {
            nodes[i].remove();
        }
        ...
      }
  \end{lstlisting}
\end{lrbox}

\newsavebox\boxCsExample
\begin{lrbox}{\boxCsExample}
  \begin{lstlisting}[language=java-pretty, numbers=left]
      public static Document ParseBodyFragment(String bodyHtml, String baseUri) {
        ...
        IList <iText.StyledXmlParser.Jsoup.Nodes.Node> nodeList = ParseFragment(bodyHtml, body, baseUri);
        iText.StyledXmlParser.Jsoup.Nodes.Node[] nodes = nodeList.ToArray(new iText.StyledXmlParser.Jsoup.Nodes.Node[nodeList.Count]);
        for (int i = nodes.Length - 1; i > nodeList.Count; i--) {
            nodes[i].Remove();
        }
        ...
      }
  \end{lstlisting}
\end{lrbox}

\newsavebox\boxCsEditTranslationExample
\begin{lrbox}{\boxCsEditTranslationExample}
  \begin{lstlisting}[language=java-pretty, numbers=left]
      ...
      - for (int i = nodes.Length - 1; i > (-w<)nodeList.Count(>w-); i--) {
      + for (int i = nodes.Length - 1; i > (+w<)0(>w+); i--) {
      ...
  \end{lstlisting}
\end{lrbox}

\newsavebox\boxCsUpdateExample
\begin{lrbox}{\boxCsUpdateExample}
  \begin{lstlisting}[language=java-pretty, numbers=left]
      ...
      - iText.StyledXmlParser.Jsoup.Nodes.Node[] nodes = nodeList.ToArray(new iText.StyledXmlParser.Jsoup.Nodes.Node[(-w<)nodeList.Count(>w-)]);
      + iText.StyledXmlParser.Jsoup.Nodes.Node[] nodes = nodeList.ToArray(new iText.StyledXmlParser.Jsoup.Nodes.Node[(+w<)0(>w+)]);
      - for (int i = nodes.Length - 1; i > (-w<)nodeList.Count(>w-); i--) {
      + for (int i = nodes.Length - 1; i > (+w<)0(>w+); i--) {
      ...
  \end{lstlisting}
\end{lrbox}

\newsavebox\boxCsTranslation
\begin{lrbox}{\boxCsTranslation}
  \begin{lstlisting}[language=java-pretty, numbers=left]
      ...
      - (-w<)iText.StyledXmlParser.Jsoup.Nodes.(>w-)Node[] nodes = nodeList.ToArray(new (-w<)iText.StyledXmlParser.Jsoup.Nodes.(>w-)Node[(-w<)nodeList.Count(>w-)]);
      + Node[] nodes = nodeList.ToArray(new Node[(+w<)0(>w+)]);
      - for (int i = nodes.Length - 1; i > (-w<)nodeList.Count(>w-); i--) {
      + for (int i = nodes.Length - 1; i > (+w<)0(>w+); i--) {
      ...
  \end{lstlisting}
\end{lrbox}

\begin{tikzpicture}[
    line width=0.4pt,
    node distance=0ex and 0em,
    every node/.style={scale=1},
    gridBox/.style={rectangle, opacity=0, draw=red},
    box/.style={rectangle, draw=black, inner sep=2pt, font=\small},
    rounded box/.style={rectangle, rounded corners, draw=black, inner sep=2pt, font=\small},
    anno/.style={font=\footnotesize},
]

    \DefMacro{wCodeBox}{46em}
    \DefMacro{hCodeBox}{19ex}
    \DefMacro{hJavaCodeBox}{23ex}
    \DefMacro{hCsCodeBox}{21ex}
    \DefMacro{hCsEditTranslationCodeBox}{10ex}
    \DefMacro{hCsUpdateCodeBox}{14ex}
    \DefMacro{hCsTranslationBox}{14ex}
    \DefMacro{ModelNameY}{-10pt}
    \DefMacro{color-box-w}{45.5em}
    
    \node (g-Old) at (0,0) [box, minimum width=\UseMacro{wCodeBox}, minimum height=\UseMacro{hJavaCodeBox}] {};
    \node (j-del) at(0, 0.1)[box, minimum width=\UseMacro{color-box-w}, minimum height=2ex, draw=none, fill=\UseMacro{git-del-color}, ] {};
    \node (j-add) at(0, -0.17)[box, minimum width=\UseMacro{color-box-w}, minimum height=2ex, draw=none, fill=\UseMacro{git-add-color}, ] {};
    \node (b-Old) [right = -0.3 of g-Old.west] [] {\usebox\boxJavaCode};
    \node (b-OldText) [below left = \UseMacro{ModelNameY} and 1pt of g-Old.south east] [box, scale=0.8] {\bf Java Change\phantom{p\hskip -.6em}};
    \node (g-New) [below = 0 of g-Old.south] [box, minimum width=\UseMacro{wCodeBox}, minimum height=\UseMacro{hCsCodeBox}] {};
    \node (b-New) [right = -0.3 of g-New.west] [] {\usebox\boxCsExample};
    \node (box-edit-translation) [below = 0 of g-New.south] [box, minimum width=\UseMacro{wCodeBox}, minimum height=\UseMacro{hCsEditTranslationCodeBox}] {};
    \node (c-del-0) at(0, -5)[box, minimum width=\UseMacro{color-box-w}, minimum height=2ex, draw=none, fill=\UseMacro{git-del-color}, ] {};
    \node (c-add-0) at(0, -5.28)[box, minimum width=\UseMacro{color-box-w}, minimum height=2ex, draw=none, fill=\UseMacro{git-add-color}, ] {};
    \node (edit-translatioin) [right = -0.3 of box-edit-translation.west] [] {\usebox\boxCsEditTranslationExample};
    \node (box-update) [below = 0 of box-edit-translation.south] [box, minimum width=\UseMacro{wCodeBox}, minimum height=\UseMacro{hCsUpdateCodeBox}] {};
    \node (c-del-1) at(0, -6.37)[box, minimum width=\UseMacro{color-box-w}, minimum height=2ex, draw=none, fill=\UseMacro{git-del-color}, ] {};
    \node (c-add-1) at(0, -6.65)[box, minimum width=\UseMacro{color-box-w}, minimum height=2ex, draw=none, fill=\UseMacro{git-add-color}, ] {};
    \node (c-del-2) at(0, -6.93)[box, minimum width=\UseMacro{color-box-w}, minimum height=2ex, draw=none, fill=\UseMacro{git-del-color}, ] {};
    \node (c-add-2) at(0, -7.21)[box, minimum width=\UseMacro{color-box-w}, minimum height=2ex, draw=none, fill=\UseMacro{git-add-color}, ] {};
    \node (update) [right = -0.3 of box-update.west] [] {\usebox\boxCsUpdateExample};
    \node (box-translation) [below = 0 of box-update.south] [box, minimum width=\UseMacro{wCodeBox}, minimum height=\UseMacro{hCsTranslationBox}] {};
    \node (c-del-3) at(0, -8.27)[box, minimum width=\UseMacro{color-box-w}, minimum height=2ex, draw=none, fill=\UseMacro{git-del-color}, ] {};
    \node (c-add-3) at(0, -8.55)[box, minimum width=\UseMacro{color-box-w}, minimum height=2ex, draw=none, fill=\UseMacro{git-add-color}, ] {};
    \node (c-del-4) at(0, -8.83)[box, minimum width=\UseMacro{color-box-w}, minimum height=2ex, draw=none, fill=\UseMacro{git-del-color}, ] {};
    \node (c-add-4) at(0, -9.11)[box, minimum width=\UseMacro{color-box-w}, minimum height=2ex, draw=none, fill=\UseMacro{git-add-color}, ] {};
    \node (translation) [right = -0.3 of box-translation.west] [] {\usebox\boxCsTranslation};
    \node (b-newtext) [below left = -9pt and 1pt of g-New.south east] [box, draw, scale=0.8] {\bf C\# Old Method};
    \node (b-newtext) [below left = -10pt and 1pt of box-edit-translation.south east] [box, draw, scale=0.8] {\bf \DeltaTool (\edittranslation) Prediction\phantom{p\hskip -.6em}};
    \node (b-newtext) [below left = \UseMacro{ModelNameY} and 1pt of box-update.south east] [box, draw, scale=0.8] {\bf \UseMacro{CodeT5Up} Prediction\phantom{p\hskip -.6em}};
    \node (b-newtext) [below left = \UseMacro{ModelNameY} and 1pt of box-translation.south east] [box, draw, scale=0.8] {\bf \translationModel Prediction\phantom{p\hskip -.6em}};
    
\end{tikzpicture}
  \vspace{-5pt}
  \caption{Qualitative analysis of all the models on one example in the test data of \UseMacro{java2cs} dataset.\label{fig:results-example}}
\end{figure*}

To further investigate the advantages of \DeltaTool over the best
generation-based model
(\UseMacro{CodeT5Up}), we break down the performance of
\edittranslation and \UseMacro{CodeT5Up} on each example in the test
data of \UseMacro{java2cs}.  In Figure~\ref{fig:xmatch-token-number},
we show the average percentage of \DeltaTool (\edittranslation) and
\UseMacro{CodeT5Up}'s predictions that exactly match the ground truth
with respect to the number of sub-tokens in the input old code
(\UseMacro{cs_old}).  Note that the code are subtokenized using the
Roberta tokenizer~\cite{Roberta}, which is used by all machine
learning models.  We exclude the examples that have more than 500
sub-tokens from being shown in this figure as those outliers only
account for less than 5\% of the test data.  We can see that the
performance of \UseMacro{CodeT5Up} drastically drops with the increase
of number of sub-tokens in the code to be edited (\UseMacro{cs_old}),
but \edittranslation's performance is rather stable.  This illustrates
another benefit of \DeltaTool in accurately handling longer input,
because of focusing on transforming the edits instead of generating
the entire new code like \UseMacro{CodeT5Up}.

Meanwhile, most of the existing transformer-based models have a length
limit for the input sequence because the naive self-attention has quadratic complexity with regard to the input length.
In Figure~\ref{fig:ref-token-number},
we present the distribution of the number of sub-tokens in the models'
target outputs for \DeltaTool (\edittranslation) and
\UseMacro{CodeT5Up} on the test data of \UseMacro{java2cs}.  We only
show the distribution of target outputs with fewer than 500 sub-tokens
for the same reason described in the previous paragraph.  Most of \DeltaTool's target outputs (the
sequence of edit operations) are shorter than \UseMacro{CodeT5Up}'s
output (new code in target programming language).  This might explain
why \DeltaTool achieves better performance than generation-based
models on longer code as generating longer sequence are generally more
challenging to machine learning models.
Recent studies~\cite{longformer, flashattention, unlimiformer} have focused on exploring approaches to address the limitation of the model's input context window size. Future research should examine the performance difference between translating edit sequences and generating entirely new code using models capable of handling longer context.

\RQ{3}{\textbf{Combining generation-based model with \DeltaTool.}}
To exploit the superiority of generation-based model on short code
snippets, we combine our strongest generation
model---\UseMacro{CodeT5Up}---with the strongest \DeltaTool
\mode---\edittranslation---based on the size of the code snippet.
Specifically, we use \UseMacro{CodeT5Up} if the code to be updated has
fewer sub-tokens than a threshold and use \DeltaTool
(\edittranslation) otherwise.
To pick the threshold for combining two models, we performed a
grid-search on the validation set and selected the one that gives
optimal xMatch score.  
 We refer to
the combined model as the \emph{Hybrid} model and provide its results
on the bottom row of Table~\ref{tab:java2cstranslation-results} to Table~\ref{tab:cs2java-cp-translation-results}.
By combining generation-based model with \DeltaTool, we can achieve
improved performance on most of the reported automatic metrics.

\subsection{Qualitative Analysis}
\label{sec:eval:qualitative}
Figure~\ref{fig:results-example} shows an example in
\UseMacro{java2cs} dataset and the models' predictions. We show the
code changes from \Java project \CodeIn{itext/itext7} in the method
(\CodeIn{parseBodyFragment}).  The newly added code is highlighted in
green and removed code is highlighted in red.  We also present the
old version of the corresponding \Cs method
(\CodeIn{ParseBody}\-\CodeIn{Fragment}) from \CodeIn{itext/itext7\text{-}dotnet},
and the predicted code changes from three models: \DeltaTool
(\edittranslation), \UseMacro{CodeT5Up}, \translationModel.  Note that
\translationModel only has access to the new version of \Java{}
method.

Although \translationModel is able to correctly translate the code
change in \Java, it fails to infer the full name of the type
\CodeIn{Node} and makes an irrelevant edit, because it does not have
the context of the old version of \Cs code.  
\UseMacro{CodeT5Up} correctly captures the \Java change while making an extra
irrelevant edit on the \Cs code.
 Our proposed model, \DeltaTool
(\edittranslation) accurately identifies the position in the \Cs
method to make edits and correctly adjusts the \Java{} edits.

\section{Limitations}

\MyPara{Studied programming languages}.
We study the translation of code changes between two programming
  languages.  In this paper, we focus on open-source \Java and \Csharp
  projects due to the ease of locating corresponding changes using
  heuristics.  Nevertheless, it is important to note that our approach
  can be applied to other programming language pairs as well, and we
  leave the investigation of such pairs for future research.

\MyPara{Correspondence between programming languages}
Our model, \DeltaTool, is intended for developers to migrate
  code changes from a project written in a source programming language
  to projects written in target programming languages, leveraging
  known correspondences (e.g., methods with similar functionalities)
  between the source and target programming languages.  In this work,
  we adopt a similar strategy used in~\cite{CodeXGLUE} to match \Java
  and \Csharp methods.  In practice, a code retrieval system can be
  used as a first step to identify the locations where the code
  changes should be propagated.  We leave the combination of code
  retrieval tool and \DeltaTool as future work.

\MyPara{Empirical evaluation}
This paper presents the empirical study results for internal
  metrics that are of interest to researchers. However, the external
  measurements of the impact on software engineering effort are not
  included in this study. These measurements could be addressed by
  conducting user studies.

\section{Related Work}

In this section, we describe related work on the rule-based code
translation tools, existing machine learning models designed for code
translation, and the machine learning models that are proposed for
accelerating software evolution.

\MyPara{Rule-based code translation}
Researchers and practitioners have designed rule-based tools for
translating the source code between programming languages.
Such tools, usually called transpilers, were built for pairs like
\Java and \Cs{}~\cite{java2csharp}, C and Rust~\cite{c2rust}, C and
Go~\cite{c2go}.  \citet{NguyenETAL15Divide} proposed PBSMT, a
phrase-based statistical machine translation models for source code
translation.
\citet{gyori2013crossing} proposed \textsc{LambdaFicator} to translate
imperative Java code to using the functional Stream APIs.
\citet{radoi2014translating} presented the rule-based
model to translate sequential \Java code to MapReduce framework.  
Prior work~\cite{mariano2022automated} has shown that existing
rule-based code refactoring tools can only deal with stylized code
snippets over common code patterns.

\MyPara{Learning-based code translation}
Researchers have proposed various machine learning models for the code
translation task.  \citet{ChenETAL18Tree} proposed a tree-to-tree
neural network with a tree-RNN encoder and a tree-RNN decoder.  
Motivated by the success of large \pretrained \LLMs for many Natural
Language Processing tasks, domain-specific models that are \pretrained
on source code and technical text have emerged.
Researchers have applied them to the code translation task.
\citet{CodeXGLUE} proposed CodeXGLUE, a benchmark including the code
translation dataset consisting of \Java and \Cs methods with
equivalent functionality.  They \finetuned and evaluated CodeBERT on the
translation dataset.
Results showed that it produced
the best results among all the existing baselines.
LLMs that are built on the encoder-decoder paradigm and \pretrained
with general unsupervised denoising auto-encoding objectives showed
promising results on wide range of code generation tasks including
code translation.  Such models include CodeT5~\cite{WangETAL21codet5},
PLBART~\cite{PLBART}, and UniXcoder~\cite{GuoETAL22Unixcoder}.  For
the comparison of \DeltaTool with state-of-the-art code translation
models, we include two variants of the CodeT5-based translation models
(with history context and without) in our evaluation.

Researchers designed LLMs which are \pretrained with the
objective tailored for code translation.
\citet{TipirneniETAL22StructCoder} introduced tasks on predicting AST
paths and data flows during \pretraining.
\citet{LachauxETAL20TransCoder} proposed TransCoder which is
\pretrained to do code translation with
back-translation objective.
To improve the quality of \pretraining data, \citet{RoziereETAL21TransCoderST} leveraged an
automated unit-testing system to filter out invalid generated programs
during back-translation.  \citet{ZhuETAL22Multilingual} proposed
MuST, which is a multilingual code snippet translation \pretraining
objective.  None of the above work leverages the code
change history, which is the main contribution of our paper.  We leave
improving \DeltaTool with \pretraining objectives tailored for code
translation as future work.

\MyPara{Software evolution and machine learning}
New research initiatives have emerged around building and evaluating
models that aid the process of software evolution.
Prior work \cite{PanthaplackelETAL20Learning, liu2021just, gao2021automating, lin2022predictive, LinETAL21CommentUpdate} proposed to update the comment
given the changes in the associated method, e.g., \citet{PanthaplackelETAL20Learning}  built a model that
takes the code change as context to make edits on the outdated
comment.  \citet{NieETAL22EvalMethodologies} present different
approaches to split dataset into training, validation and test sets
and studied how different approaches affect the evaluation of machine
learning models.  
\citet{kamezawaETAL22RNSum} presented a dataset, RNSum, which consists
of release notes and the associated commit messages
collected from GitHub repositories and designed models to generate
release notes based on the commit messages.
\citet{ZhangETAL22CoditT5} proposed a novel \pretraining objective
designed for software editing tasks and built CoditT5.  CoditT5 was \finetuned on three downstream tasks
related to the software evolution.
\citet{LiETAL2022Automating, tufano2022using, ZhangETAL22Using}
proposed models that targeted various tasks through the code review
process.
The models are trained on the historical data and evaluated on the new
pull requests submitted for code review.  Our \DeltaTool model
incorporates the context from the code changes in source programming
language and the old version of method in target programming languages
to improve its performance on the \deltatask task, which helps
developers co-evolve the projects implemented in different programming
languages.

\section{Conclusion}
In this paper, we formulated a new task: translating code changes
across programming languages with the goal to synchronize projects
that provide the same \APIs or implementations in multiple programming
languages.
We proposed \DeltaTool, a model which uses code change history as
contextual information and learns to make edits on the existing
version of code written in the target programming language.
We showed that our model outperforms existing code translation models
and is better than the generation-based models even if they use
historical context.
\DeltaTool is a significant advancement in supporting developers with
the maintenance of their projects that incrementally provide identical
functionalities in multiple programming languages.

\section*{Acknowledgments}
We thank Nader Al Awar, Yu Liu, Sheena Panthaplackel, Aditya
Thimmaiah, Zhiqiang Zang, and the anonymous reviewers for their
comments and feedback.
We acknowledge the Texas Advanced Computing Center (TACC) at The
University of Texas at Austin for providing HPC resources that have
contributed to the research results reported within this paper.
This work is partially supported by the US National Science Foundation
under Grant Nos.~CCF-2107291, IIS-2145479, CCF-2217696 and CCF-2313027.

\bibliography{bib}
\balance

\end{document}